%% file: SBHSarxiv.tex
\documentclass[12pt,a4]{article}

\usepackage{amsthm,amsmath,latexsym,amssymb,amsfonts,amscd}
\usepackage{graphics,lscape,fancyhdr,array,stmaryrd,euscript}
\usepackage{empheq}
\usepackage{dsfont}
\usepackage{verbatim}
\usepackage{color,tikz,tikz-cd}
\usetikzlibrary[snakes]
\usepackage{relsize,slashed}
\numberwithin{equation}{section}
\usepackage{hyperref}
\input{YoungDiagrams.tex}

\newcommand{\pl}{\partial}

\usepackage[all]{xy}
\usepackage{hyperref}

\usepackage[numbers,sort&compress]{natbib}
\usepackage[nottoc]{tocbibind}

\newcolumntype{w}[1]{%
>{\raggedright\hspace{0pt}}m{#1}}%
\newcolumntype{z}[1]{%
>{\raggedleft\hspace{0pt}}m{#1}}%

\newcommand{\besubeqs}{\begin{subequations}}
\newcommand{\esubeqs}{\end{subequations}}

\newcommand{\trA}{{\mathrm{Tr}}}

\newcommand{\Tr}[1]{{\left\langle #1 \right\rangle}}
\newcommand{\Trace}[2]{{\mathrm{Tr}_{#1}\!\left[ #2 \right]}}
\newcommand{\inv}{{\mathtt{I}}}

\newcommand{\hs}{{\ensuremath{\mathfrak{hs}}}}
\newcommand{\prd}{\diamond}
\newcommand{\hsdouble}{{\ensuremath{{\mathfrak{hs}}_0}}}

\newcommand{\wv}{{\mathbb{J}}}
\newcommand{\up}{\nu}
\newcommand{\upA}{\nu_+}
\newcommand{\upB}{\nu_-}

\newcommand{\gers}[1]{{\llbracket #1\rrbracket}}

\begin{document}
\pagenumbering{gobble}
\hfill
\vskip 0.01\textheight
\begin{center}
{\Large\bfseries 
Slightly Broken Higher Spin Symmetry: \\
\vspace{0.2cm}
General Structure of Correlators
\vspace{0.4cm}}

\vskip 0.03\textheight
\renewcommand{\thefootnote}{\fnsymbol{footnote}}
Pavel \textsc{Gerasimenko}${}^{a,d}$, Alexey \textsc{Sharapov}${}^{b}$ \&  Evgeny \textsc{Skvortsov}\footnote{Research Associate of the Fund for Scientific Research -- FNRS, Belgium}${}^{c,d}$
\renewcommand{\thefootnote}{\arabic{footnote}}
\vskip 0.03\textheight

{\em ${}^{a}$Department of General and Applied Physics, \\
Moscow Institute of Physics and Technology, \\
Institutskiy per. 7, Dolgoprudnyi, \\141700 Moscow region, Russia}

{\em ${}^{b}$Physics Faculty, Tomsk State University, \\Lenin ave. 36, Tomsk 634050, Russia}\\
\vspace*{5pt}
{\em ${}^{c}$ Service de Physique de l'Univers, Champs et Gravitation, \\ Universit\'e de Mons, 20 place du Parc, 7000 Mons, 
Belgium}\\
\vspace*{5pt}
{\em ${}^{d}$ Lebedev Institute of Physics, \\
Leninsky ave. 53, 119991 Moscow, Russia}\\

\end{center}

\vskip 0.02\textheight

\begin{abstract}
We explore a class of CFT's with higher spin currents and charges. Away from the free or $N=\infty$ limit the non-conservation of currents is governed by operators built out of the currents themselves, which deforms the algebra of charges by, and together with, its action on the currents. This structure is encoded in a certain $A_\infty/L_\infty$-algebra. Under quite general assumptions we construct invariants of the deformed higher spin symmetry, which are candidate correlation functions. In particular, we show that there is a finite number of independent structures at the $n$-point level. The invariants are found to have a form reminiscent of a one-loop exact theory. In the case of Chern--Simons vector models the uniqueness of the invariants implies the three-dimensional bosonization duality in the large-$N$ limit. 
\end{abstract}
\newpage
\tableofcontents
\newpage
\section{Introduction}
\pagenumbering{arabic}
\setcounter{page}{1}
Infinite-dimensional symmetries is a kingly gift in many interesting models of quantum field theory, which makes them tractable if not soluble. Virasoro symmetry is one of the simplest examples of an ordinary Lie-type symmetry that is powerful enough to solve some models of $2d$ conformal field theories \cite{Belavin:1984vu}. Another example of a completely different nature is Yangian symmetry, which is a Hopf algebra governing the planar $\mathcal{N}=4$ SYM \cite{Beisert:2006fmy,Drummond:2009fd,Beisert:2010jr}. Strong homotopy algebras of $L_\infty$ and $A_\infty$ types constitute a further class of mathematical structures that extend the conventional notion of symmetry \cite{Zwiebach:1992ie,Gaberdiel:1997ia,Kajiura:2003ax,Lada:1992wc,Alexandrov:1995kv,Barnich:2004cr,Hohm:2017pnh}. These are quite general and have nothing to do with integrability or solubility {\it per se}. In physics, BV-BRST formalism is closely related to $L_\infty$-algebras with an important additional constraint due to locality in QFT. In general, $A_\infty/L_\infty$-algebra is a way to encode the formal consistency of dynamics, but it can in some circumstances be responsible for integrability as well. 

Our interest is yet another infinite-dimensional symmetry -- a higher spin symmetry, which manifests itself in conserved higher rank tensors, $\pl^m J_{ma_2...a_s}=0$. This symmetry is clearly present in free theories and disappears in interacting ones \cite{Zilch, Deser:1980fk,Maldacena:2011jn,Boulanger:2013zza,Alba:2013yda,Alba:2015upa}. An important observation made in \cite{Maldacena:2012sf} is that in certain models, e.g. (Chern--Simons) vector models in $3d$, the higher spin symmetry gets deformed in a very constrained way rather than broken. This was dubbed the slightly-broken higher spin symmetry. The deformation can be understood as an $A_\infty/L_\infty$-algebra that deforms a given higher spin algebra in a certain way \cite{Sharapov:2018kjz}. The goal of this paper is to construct invariants of the slightly-broken higher spin symmetry, which should compute the correlation functions. The uniqueness of these invariants in the case of Chern--Simons matter theories \cite{Sharapov:2020quq} implies the three-dimensional bosonization duality \cite{Giombi:2011kc, Maldacena:2012sf, Aharony:2012nh,Aharony:2015mjs,Karch:2016sxi,Seiberg:2016gmd} in the large-$N$ limit. 

The unbroken higher spin symmetry can be summarized by saying that (i) the higher spin currents $J_s$, with $s=2$ member being the stress-tensor, generate the higher spin charges $Q_s$; (ii) $Q_s$ have to form a Lie algebra $\hs$, the higher spin algebra, which is infinite-dimensional and contains the conformal algebra $so(d,2)$; (iii) currents $J$ form a module over $\hs$:
\begin{align}
    \pl\cdot J_s&=0 && \Longrightarrow && Q_s=\textstyle\int J_s && \Longrightarrow && [Q,Q]=Q\quad \&\quad [Q,J]=J\,.
\end{align}
A crucial property of every free CFT is that $\hs$ originates from an associative algebra and the Lie bracket is given by the commutator. This associative algebra encodes all important information about a given CFT.  
It then comes as no surprise that each higher spin algebra admits a simple set of invariants (traces) $\Trace{\star}{\wv\star ...\star \wv}$, which are automatically conformally-invariant since $so(d,2)\subset \hs$. For an appropriate choice of wave functions $\wv$, these invariants are nothing but the correlation functions of $J$'s and can be computed explicitly, ending up with manifestly conformally-invariant expressions \cite{Colombo:2012jx,Didenko:2012tv,Didenko:2013bj, Bonezzi:2017vha}. The complete generating function is given by the logarithm  of the determinant
\begin{align}
    W_{\text{free}}[\wv]&=\trA_\star\log_\star [1-\wv]\,.
\end{align}
Indeed, $J_s$ are bilinear in the fundamental fields, e.g. $J_s=\phi \pl...\pl \phi+...$, and the generating functional of the correlators is a Gaussian path integral. This expression is obvious for a free theory, but is also true for the higher spin currents in the critical vector models at $N=\infty$. With $V$ denoting the $so(d,2)$-module where the fundamental field(s) take values, it is clear that $J$'s together with their descendants span the space $V\otimes V$ and $\hs$ is isomorphic to  $\mathrm{End}(V)\sim V\otimes V^*$, which explains why $\hs$ is associative and acts naturally on both $\phi$'s and $J$'s.

When we depart from the free or $N=\infty$ limit the interactions break the conservation of the higher spin currents. The stress-tensor has to remain conserved, of course. The conservation turns into a non-conservation. In general one would conclude that the symmetry is gone and is no longer useful \cite{Deser:1980fk}, which is indeed the case if the non-conservation is driven by some other operators. A remarkable  feature of the higher spin symmetry breaking in vector models is that the non-conservation of higher spin currents is still driven by the higher spin currents themselves, by the double-trace, $[JJ]$, (and triple-trace, sometimes) operators that are built out of $J$'s. Since the currents are no longer conserved the charges $Q$ cannot form a Lie algebra. In addition, the action of $Q$ on $J$ is deformed as well:  
\begin{align*}
    \pl\cdot J&=g\,[JJ] && \Longrightarrow && Q=\textstyle\int J && \Longrightarrow && [Q,Q]=Q+\mathcal{O}(g)\quad \&\quad [Q,J]=J+g[JJ]\,,
\end{align*}
i.e. currents generate charges whose algebra and action is deformed due to the non-conservation governed by the currents themselves.  This allows one to close the loop and to apply the idea of bootstrap. Therefore, the higher spin symmetry gets deformed rather than broken and can still be very useful. One can try to explore the non-conservation of higher spin currents directly by writing down the most general ansatz for the non-conservation equation, correlation functions etc. \cite{Maldacena:2012sf}. However, it is quite difficult to extract information this way, i.e. without understanding what happens to the higher spin symmetry. If we want to deform a Lie algebra together with its action on a module then an appropriate mathematical structure is $L_\infty$-algebra. With parameters of the higher spin symmetry denoted $\xi$, one should be looking for the deformed action $\delta_\xi$ on currents $J$ such that 
\begin{align}
   \delta_\xi J&= l_2(\xi,J)+g\,l_3(\xi,J,J)+...\,, & [\delta_{\xi_1},\delta_{\xi_2}]&=\delta_{\xi}\,, & \xi =l_2(\xi_1,\xi_2)+g\,l_3(\xi_1,\xi_2,J)+...\,, 
\end{align}
where the structure maps $l_k$ form an $L_\infty$-algebra. This seems  to be the only way to get a consistent deformation. A systematic procedure of constructing such an $L_\infty$-algebra for any given higher spin algebra was proposed in \cite{Sharapov:2018kjz}. The next step, and it is what the present paper is concerned with,  is to look for invariants of the deformed higher spin symmetry, which would deform the free CFT's correlation functions.

It is tempting to think of the higher spin symmetry as a proper replacement of the Virasoro symmetry in the case of vector models, but the way it works is completely different. Both symmetries contain the conformal algebra $so(d,2)$ and are infinite-dimensional which is what makes them efficient. However, Virasoro symmetry is a proper Lie symmetry, while higher spin symmetry gets deformed into an $L_\infty$-algebra. As a result, many applications of Virasoro rely on its representation theory \cite{Belavin:1984vu}, while $L_\infty$ has to do with the higher cohomology of higher spin algebras. Applications require Chevalley--Eilenberg cohomology, which can be reduced to the cyclic and then to the Hochschild cohomology of higher spin algebras as associative algebras \cite{Sharapov:2020quq}. The latter is the first important property of higher spin symmetry -- it originates from associative algebras, which eventually governs all of the Lie-type structures, e.g. $L_\infty$ originates from a certain $A_\infty$. The second important property is that all of the $A_\infty/L_\infty$-structures originate from another associative algebra $\hs_{\up}$, which is a certain deformation of the initial higher spin algebra $\hs$. All problems can be reduced to the theory of this deformed higher spin algebra, which is just an associative algebra.

In this paper, we make a number of structural observations regarding the slightly-broken higher spin symmetry. In particular, we prove existence and explicitly construct invariants of the deformed symmetry, which are the candidate correlation functions. These invariants are known to give free CFT correlators at the leading order \cite{Colombo:2012jx,Didenko:2012tv,Didenko:2013bj, Bonezzi:2017vha}. The results of the paper are more general and apply to any system whose symmetry gets deformed following the vector models' pattern. As it will be made clear below, any one-parameter family of associative algebras $\hs_{\up}$ leads to $A_\infty/L_\infty$-algebras and to the associated invariants. Therefore, the construction of invariants in the present paper covers this most general case as well. 

The main result of the paper is the construction of the invariants of the deformed higher spin symmetry, i.e. of the corresponding $L_\infty$-algebra, which smoothly deform the free CFT correlation functions. Surprisingly, with some abuse of notation the generating function can be written in a suggestive way
\begin{align}\label{I1}
    W[\wv]&=\trA_\prd\,{\log_\prd [1-\wv]} \, \mod \, \text{irrelevant}\,,
\end{align}
which seems to originate from a quasi-free theory. Here, $\prd$ and $\trA_\prd$ are the product and the trace in the deformed higher spin algebra $\hs_{\up}$. It is important that certain terms in the formal expansion (\ref{I1}) need to be dropped for the expression to be correct. 

Since we do not refer to any particular microscopical model that realizes a deformed higher spin symmetry, our conclusions, e.g. the form of the correlation functions, apply to all (dual) realizations at once. In the presence of a singlet constraint the correlation functions of $J_s$ encode information about all the other singlet operators in the large-$N$ limit. Both the deformation $\hs_\up$ of the algebra $\hs$ and of the trace may depend on several free parameters, which are determined by the size of the corresponding Hochschild cohomology group. These parameters are the phenomenological parameters, which can be related to the microscopical ones, if needed. By construction, $\prd$ and $\trA_\prd$ depend analytically on these parameters.  
Therefore, an important prediction for any system with the same higher spin symmetry deformation pattern, which follows from this result, is that there is a finite number of different (conformal) structures for $n$-point functions and the phenomenological coupling constants can appear only in front of them (i.e. the functions of cross-ratios cannot have a nontrivial dependence on the effective coupling constants).  

More specifically, one of the most interesting examples of the deformed higher spin symmetry is observed in Chern--Simons matter theories, see e.g. \cite{Giombi:2011kc, Maldacena:2012sf, Giombi:2016zwa,Jain:2020puw}. Some interesting facts that have been known include: (i) the higher spin algebras of free boson and free fermion are the same in $3d$, which is a necessary condition for the bosonization to take place; (ii) the three-point functions are fixed by the deformed higher spin symmetry \cite{Maldacena:2012sf}, see \cite{Giombi:2016zwa} for most and \cite{Skvortsov:2018uru} for all of them; (iii) in \cite{Sharapov:2020quq}, a complete classification of higher spin invariants was obtained. It was shown that the deformed higher spin symmetry has exactly one invariant to serve as an $n$-point function and this invariant is unobstructed. For the simplest case of vector models with higher spin currents of even spins, the deformation depends on two phenomenological parameters.

Together with \cite{Sharapov:2020quq} the results of the present paper strongly indicate the three-dimensional bosonization duality: every CFT with a slightly-broken higher spin symmetry (or any other theory with the same type of symmetry) has to have these correlations functions, irrespective of its microscopical realization (whether it is given by Chern--Simons plus bosonic or fermionic matter). For our results to apply to the widest possible class of models, we do not go into specific details. What this paper is missing at present is an explicit form of the correlation functions in terms of coordinates and polarization vectors, which will be given elsewhere.

The outline of the paper is as follows. In Section \ref{sec:unbroken}, we review the main facts about unbroken/exact higher spin symmetry to stress that correlators of $J_s$ are invariants of the higher spin symmetry. In Section \ref{sec:broken}, we review the construction of the strong homotopy algebra that deforms any given higher spin algebra \cite{Sharapov:2018kjz}. Section \ref{sec:correlators} is devoted to the invariants of the deformed higher spin symmetry. Our conclusions and discussion are in the last section.

\section{Unbroken Higher Spin Symmetry}
\label{sec:unbroken}
It is useful to start from the case of a free CFT in $d\geq3$ where higher spin symmetry is realized as an ordinary global symmetry. As we go on we will need less and less details about its structure and eventually everything will be reduced to having an associative algebra. So, our slogan is 
$$
\mbox{\bf Free CFT = Associative Algebra}\,.
$$
It  should be noted, however, that the machinery of non-commutative geometry we employ here works for more general associative algebras than those originating from free CFT's.

The stock of free CFT's is quite big, especially if we abandon unitarity: free scalar $\square \phi=0$; free fermion $\slashed \pl \psi=0$; $d/2$-forms like $F_{\mu\nu}$ in $d=4$; higher derivative theories $\square^k \phi=0$ and $\square^{k-1}\slashed \pl\psi=0$ are also conformally invariant; any other conformally-invariant equation would work. In addition, one can combine various matter options, add supersymmetry, add usual finite-dimensional global symmetries like $U(N)$, $O(N)$, weakly gauge some of them etc. For concreteness wherever needed we will bear in mind an example of the free scalar CFT.

Every free CFT's spectrum contains infinitely many conserved tensors on top of the stress-tensor, $\pl^m J_{ma_2...a_s}=0$, which are loosely called higher spin currents, see e.g. \cite{Craigie:1983fb,Anselmi:1999bb}.\footnote{The opposite statement is also true: any CFT in $d\geq3$ with at least one higher spin current has infinitely many of them and is a free CFT in disguise \cite{Maldacena:2011jn,Boulanger:2013zza,Alba:2013yda,Alba:2015upa}. Higher spin currents are not always symmetric tensors, tensors of more general symmetry are allowed. For example, they are present in the free Maxwell or fermion CFT's \cite{Alkalaev:2012rg}. All these cases are covered automatically once we arrive at an abstract associative algebra below. } These tensors are bilinear in the fundamental fields of a given free CFT, e.g. $J_{a_1... a_s}=\phi\pl_{a_1}...\pl_{a_s}\phi+...$. Last but not least, critical vector models in the $N=\infty$ limit also have higher spin currents and to the leading order in $1/N$ correlators $\langle J\ldots J\rangle$ are given by the same free Wick contractions leading to one-loop diagrams. 

The conserved tensors are indicative of an extension of the conformal symmetry. As with the conformal symmetry, the extension is due to the usual Noether symmetry: the conserved currents can be constructed by contracting the conserved tensors with (conformal) Killing tensors:
\begin{align}
    j_m(v)&= J_{ma_1...a_{s-1}}\, v^{a_1...a_{s-1}}\,.
\end{align}
Here $v^{a_1...a_{s-1}}(x)$ is a totally symmetric, traceless tensor obeying the condition
\begin{align}
    \pl^{(a_1} v^{a_2...a_s)}-\text{traces}=0\,,
\end{align}
which is a generalization of conformal Killing vector's equation \cite{Walker:1970un, Nikitin1991}. At this point the stress-tensor, $J_{ab}$, plays no special role and is just one of the members of the higher spin currents' multiplet. Conformal Killing tensors are parameterized by finite-dimensional representations of conformal algebra $so(d,2)$. In the simplest case of totally symmetric Killing tensors the corresponding parameters are irreducible tensors of $so(d,2)$ with the symmetry of two-row rectangular Young diagrams, see e.g. \cite{Eastwood:2002su} and refs therein:
\begin{align}
    \YoungpAA\,,\quad \YoungpBB\,,\quad \YoungpCC\,,...\quad \parbox{40pt}{\RectBRow{4}{4}{$k$}{}} \quad\Longleftrightarrow\quad \xi^{A_1...A_{k},B_1...B_{k}}\,.
\end{align}
These parameters will span the basis of a higher spin algebra. The first diagram corresponds to an anti-symmetric rank-two tensor, $\xi^{A,B}=-\xi^{B,A}$, i.e. to $so(d,2)$ itself. 
The Killing tensors are most nicely written with the help of the ambient space approach \cite{Eastwood:2002su} where the spacetime is identified with the projective cone: $X^AX_A=0$, $X^A\sim \lambda X^A$, $\lambda\neq0$. Here $X^A=(X^+,X^-,X^a)$, the metric is $\eta_{+-}=\eta_{-+}=1$ and $\eta_{ab}$ is the usual $so(d-1,1)$ invariant tensor. The relation between the higher spin algebra parameters and the Killing tensors reads \cite{Nikitin1991,Eastwood:2002su}
\begin{align}
    v^{a_1...a_{k}}(x) &= \xi^{A_1...A_{k},B_1...B_{k}}\, X_{A_1}...X_{A_{k}}\, E^{a_1}_{B_1}...E^{a_{k}}_{B_{k}}\,,
\end{align}
where $X^A=(1,-x^2/2,x^a)$ and $E^B_m= \pl_m X^B$. 
Higher spin currents $j(\xi)=j(v(\xi))$ lead to charges $Q(\xi)$ that act on all operators of a given CFT. In particular, $[Q,\phi]=\delta_\xi \phi$, where $\delta_\xi \phi=v^{a_1...a_{k}}\pl_{a_1}...\pl_{a_{k}}\phi+...$ \cite{Nikitin1991,Eastwood:2002su}. For the conformal symmetries we have
\begin{align}\label{usualConformal}
   \xi^{A,B}&:& \delta_\xi \phi&= v^m \pl_m \phi +\frac{\Delta}{d} (\pl_m v^m) \phi\,, && v^m=\xi^{A,B}X_A^{\vphantom{m}}E_B^m\,,
\end{align}
where for the free scalar $\Delta=\tfrac{d-2}{2}$ and $v^m$ is a conformal Killing vector, which is parameterized by $\xi^{A,B}$ in the adjoint of $so(d,2)$. The simplest example of a genuine higher spin symmetry is $\delta_\xi \phi=\xi^{a_1...a_{k},-...-}\pl_{a_1}...\pl_{a_{k}} \phi$. The conformal and higher symmetries taken together generate a higher spin algebra, which we denote by $\hs$. So far, it is only a Lie algebra. 

A somewhat trivial point, which will be of crucial importance later, is that the higher spin symmetries form an associative algebra rather than just a Lie algebra \cite{Nikitin1991,Eastwood:2002su}. This fact cannot be a consequence of/related to Noether's theorem as such. Indeed, given linear equations of motion $E\phi=0$, a symmetry $S$ is a differential operator that maps solutions to solutions. This implies that $ES\phi= L_S E\phi$ for some other operator $L_S$. It then follows immediately that the product $S_1S_2$ of two symmetries $S_{1}$ and $S_2$ is a symmetry again. Therefore, symmetries of linear equations form associative algebras. In our case of a free CFT, it is clear that the algebra is infinite-dimensional and  contains the conformal algebra $so(d,2)$ as Lie subalgebra. For the free scalar CFT the algebra is generated by the operators \eqref{usualConformal}, see \cite{Nikitin1991,Eastwood:2002su} and closely related \cite{Konstein:2000bi}. Hence, $\hs$ is associative and the usual Lie symmetries result from taking the commutator as Lie bracket. It is then of no surprise that $\hs$ can be obtained through the deformation quantization of a certain coadjoint orbit of $so(d,2)$.

The same higher spin algebra can be understood directly in terms of the physical states: it is the endomorphism algebra of one-particle states. For example, let us take $\phi(x)$ and construct the lowest weight state $|\phi\rangle= \phi(0)|0\rangle$. It obeys $D |\phi\rangle= \tfrac{d-2}{2}|\phi\rangle$, $K_a |\phi\rangle=0$, $L_{ab}|\phi\rangle=0$. The descendants are generated by $P_a$ and the space of one-particle states $V$ is the span of $P_{a_1}...P_{a_k}|\phi\rangle$. It is clear that the conformal algebra generators $P_a,D,L_{ab}, K_a$ act on $V$. Hence, the entire universal enveloping algebra $U(so(d,2))$ acts on $V$. This action is not free and there are many relations. For example, the Casimir operators of $so(d,2)$ acquire fixed numerical values on $V$. Also, for the scalar field we have $P_mP^m=0$. All these relations generate a two-sided ideal $I$ in $U(so(d,2))$, which in this case is called Joseph ideal. The higher spin algebra $\hs$ is given \cite{Eastwood:2002su} by the quotient $U(so(d,2))/I$, which is equivalent to $\mathrm{End}(V)\sim V\otimes V^*$.

With the last definition one can elaborate the action of the higher spin algebra on $\phi(x)$ by noting that $\phi(x)=e^{x\cdot P}\phi(0)e^{-x\cdot P}$ and, then, using the translation operator $e^{x\cdot P}$ to drag the action of any polynomial $f(P,L,K,D)$ in the conformal algebra generators to point $x$, which is how the action of the conformal algebra on any conformal operator $O(x)$ is worked out. 

One way or another, given any free CFT we arrive at an infinite-dimensional associative algebra $\hs$ that contains $so(d,2)$ as a Lie subalgebra. In all known cases this algebra has several realizations that can be helpful (e.g. quasi-conformal realization, oscillator realization, deformation quantization of the coadjoint orbit that corresponds to a given free field as a representation of $so(d,2)$, see \cite{Dirac:1963ta, Gunaydin:1981yq,Gunaydin:1983yj,Fradkin:1986ka,Vasiliev:1986qx, Gunaydin:1989um,Nikitin1991,Eastwood:2002su,Michel, Joung:2014qya,Gunaydin:2016bqx}). Any higher spin algebra $\hs$ admits a unique trace, $\Tr{\bullet}: \hs \rightarrow \mathbb{R}$, $\trA\,{[a\star b-b\star a]}=0$, which projects everything onto the unit of the algebra.

Coming to the correlation functions, in the case of the free boson CFT there is a simple generating function of conserved tensors
\begin{align}
    j\!\!j(x,y)&= \bar\phi(x-y)\phi(x+y)= \sum_s j\!\!j_{a_1...a_s}(x) y^{a_1}...y^{a_s}\,, && \pl^m j\!\!j_{ma_2...a_s}=0\,.
\end{align}
Note that $j\!\!j_s$ are not traceless and, for that reason, are not quasi-primary. Nevertheless, they are very handy. Simple Wick's contractions give the correlation functions thereof. Equivalently, the Gaussian path integral 
\begin{align}\label{CD}
    \exp W[B]&= \int D\phi\, \exp\left[- \int dx\, \bar\phi(-\pl^2) \phi + \int dx\,dy\, j\!\!j(x,y) B(y,x) \right]\,,
\end{align}
with $B(x,y)$ being the source for $j\!\!j(x,y)$, leads to $W[B]=-\mathrm{Tr}\log[1-G \cdot B]$; here $G=(-\pl^2)^{-1}\sim |x|^{-(d-2)}$ is the Green function and $\cdot$ is a shorthand for the convolution, e.g. the source term here-above is $\mathrm{Tr}[j\!\!j \cdot B]$. Partition function \eqref{CD} is a starting point for the collective dipole approach \cite{deMelloKoch:2018ivk,Aharony:2020omh}. While $W[B]$ does the job of encoding all correlation functions of $j\!\!j$ and, after a certain transformation/projection, of the higher spin currents $J_s$, it does not lead to manifestly conformally-invariant expressions, nor is the higher spin symmetry obvious \cite{David:2020ptn}.

The same free CFT correlation functions can be rewritten in a manifestly conformally and higher spin invariant way. To this end, let us recall that the
higher spin conserved tensors $J_s$ belong to $V\otimes V$, which is induced from the lowest weight state $|\phi\rangle \otimes |\phi\rangle$. Indeed, $J_s$ are single-trace operators and have to reside in the OPE $\phi(x)\phi(0)=...$.  For a free CFT, this is equivalent to taking the tensor product $V\otimes V$ and decomposing it into $so(d,2)$ modules, the only difference between OPE and the tensor product being the first term $\langle\phi(x)\phi(0)\rangle$ in the OPE. The Hermitian conjugation allows us to map $|\phi\rangle \otimes |\phi\rangle$ to $|\phi\rangle \otimes \langle \phi|$, that is, $V\otimes V$ to $V\otimes V^*$. The latter tensor product is isomorphic to the higher spin algebra $\hs$, at least formally.\footnote{A finite-dimensional analogy is that $V\otimes V$ is (non-canonically) isomorphic to $V\otimes V^\star$ and the latter is just $gl(V)$. The fundamental field $\phi$ and its descendants form $V$, the higher spin tensors form $V\otimes V$, which is formally isomorphic to the `space of matrices' $gl(V)$ on $V$, that is, to the higher spin algebra. } This observation allows one to take advantage of the trace on $\hs$ and construct the invariants
\begin{align}\label{freeTr}
    I_n[\wv] &= \tfrac1{n} \trA_\star[\wv\star ...\star \wv]\,, & W_{\hs}[\wv]&=\trA_\star\,{\log_\star [1-\wv]}=\sum_n I_n[\wv] \,,
\end{align}
together with a generating function $W[\wv]$, which is reminiscent of $W[B]$. Here, $\wv$ are some wave-functions that transform in the adjoint of $\hs$. Higher spin tensors $J$ transform according to their origin from $V\otimes V$. The Hermitian conjugation, once realized through the inversion transformation 
$\inv$, 
\begin{equation}\label{invertion}
    K^a=\inv P^a\inv \,, \qquad L^{ab}=\inv L^{ab}\inv \,, \qquad P^a=\inv K^a\inv\,, \qquad -D=\inv D\inv\,,
\end{equation} 
allows one to map $J$ to $\hs$ via $J \inv$. As a result, $J \inv$ transforms in the adjoint representation  or $J$ has the right action twisted by $\inv$, $\delta_\xi J=\xi\star J-J \star \inv(\xi)$. A useful trick here \cite{Iazeolla:2008ix} is to extend $\hs$ with a new generator $\inv$ by forming the smash product $\hsdouble=\hs \rtimes \mathbb{Z}_2$, where $\mathbb{Z}_2=(1,\inv)$. The elements of the extended algebra $\hsdouble$ are of the form $a=a'+a''\inv$ and the product is defined as
\begin{equation}
(a'+a''\inv)(b'+b''\inv)=(a'b'+a'' \inv(b'')) +(a'b''+a'' \inv(b'))\inv\,.
\end{equation} 
Here $\inv(a)=\inv a\inv $ denotes the action of the inversion on the algebra elements, which is obtained by extending (\ref{invertion}) to functions in $P^a,K^a, L^{ab}$,  and $D$. Now, both the adjoint and twisted actions are parts of the adjoint action in the extended algebra $\hsdouble$. 

Provided that wave-functions $\wv$ are chosen appropriately to represent higher spin tensors $J$ in flat space, $I_n$ compute the correlation functions of $J_s$. All $n$-point correlators of $J_s$ of the free $3d$ boson and of the free $3d$ fermion CFT's were computed this way in \cite{Colombo:2012jx,Didenko:2012tv,Didenko:2013bj, Bonezzi:2017vha},\footnote{In this context it is worth pointing out a precursor in \cite{Giombi:2010vg}, where the same three-point functions resulted from a certain regularization of a divergent bulk expression. } which proves the concept.\footnote{\label{normalization}Note that for each $n$ the higher spin symmetry fixes all $\langle J_{s_1}...J_{s_n}\rangle$. It does not, however, fix the relative coefficient between $n$-point and $(n+1)$-point correlators. Nevertheless, these relative coefficients do not have any physical significance; one can fix them by comparing with OPE's. For simplicity, one can just compute correlators of $J_0$ by Wick contractions to find these relative coefficients. It is also worth stressing that, while higher spin algebra is generated by the charges associated with $J_s$, $s>0$, all $J_s$, $s\geq0$ belong to a single representation of the higher spin algebra. In particular, $I_n$ compute correlators of all $J_s$, $s\geq0$.} Despite the holographic context of \cite{Colombo:2012jx,Didenko:2012tv,Didenko:2013bj, Bonezzi:2017vha} the computation itself has nothing to do with holography, in particular, the dependence of $\wv$ on $AdS_4$ coordinates is artificial and vanishes from the invariants. 

In general, to get $I_n$ one needs to compute the star-product (since $\hs$ can be obtained via deformation quantization) of certain wave-functions $\wv$ representing $J_s$ in the higher spin algebra, which might be tedious.\footnote{The wave-functions are extremal projectors \cite{Didenko:2012tv}, which implies the Wick theorem for correlators of $J_s$.} For the case of the $3d$ free boson/fermion CFT's, the higher spin algebra is just the even subalgebra $A_2^e$ of the Weyl algebra $A_2$ \cite{Dirac:1963ta, Gunaydin:1981yq,Gunaydin:1983yj}. This is given by even functions $f(\hat{a},\hat{a}^\dag)$ in two pairs of creation/annihilation operators $[\hat{a}^i, \hat{a}^\dag_j]=\delta^i_j$, $i,j=1,2$. Indeed, the lowest weight state $|\phi\rangle$ for free scalar (resp.  $|\psi\rangle$ for free fermion) corresponds to the Fock vacuum $\hat{a}^i|0\rangle=0$ (resp. the first exited states $\hat{a}_i^\dag|0\rangle$). The corresponding representation space $V$ is spanned by even (odd) state vectors  $f(\hat{a}^\dag_i)|0\rangle$, respectively. The relation to the spacetime picture is that $P_m\sigma^m_{ij}$ acts as $\hat{a}^\dag_i\hat{a}^\dag_j$, where $\sigma^m_{ij}$ are the Pauli matrices. 

It is now clear that elements of the higher spin algebra $\hs$ that map one-particle states to themselves are even operators $F(\hat{a}^i, \hat{a}^\dag_j)=F(-\hat{a}^i, -\hat{a}^\dag_j)$. The Weyl algebra $A_2$ can be understood as resulting from the deformation quantization of $4$-dimensional phase space, the corresponding star-product being  the Moyal-Weyl product.  This interpretation reduces the computation of correlators to simple Gaussian integrals \cite{Colombo:2012jx,Didenko:2012tv,Didenko:2013bj, Bonezzi:2017vha}, cf. \cite{Sleight:2016dba}. It is vital for the three-dimensional bosonization to work, at least in the large-$N$ limit, that the higher spin algebras of free scalar and fermion CFT's are isomorphic to each other; hence, they should  lead to the same $A_\infty$- and $L_\infty$-algebras and to the same invariants thereof. It is significant that in the strict large-$N$ limit the critical vector model and Gross--Neveu model feature the same higher spin currents as their free limits (in particular, the correlators are given by the same one-loop diagrams in momentum space). The last fact means that they also lead to the same higher spin algebras, even though they are not free CFT's.

Since Weyl algebra is ubiquitous in physics, one should not be surprised that the smash product $\hsdouble=\hs \rtimes \mathbb{Z}_2$ is already known as para-bose oscillators \cite{Luders,Druehl:1970fz,Schmutz,Ohnuki1982,Ohnuki:1984gz}. After a simple linear transformation we get $\hsdouble\sim(\mathcal{A}_0\otimes\mathcal{A}_0)/\mathbb{Z}_2$, where ($K$ is known as Klein operator \cite{Luders,Druehl:1970fz,Schmutz,Ohnuki1982,Ohnuki:1984gz,Vasiliev:1986qx})
\begin{align}
  \mathcal{A}_0\ni f(q,p,K)&: &  [q, p]&= i\,, & KqK&=-q\,,& KpK&=-p\,, & K^2&=1\,,
\end{align}
and the quotient by $\mathbb{Z}_2$ is to restrict to even functions.

Finally, it is worth making a few comments. Firstly, let us note that the higher spin Ward identities are nicely mapped to the manifest invariance of the traces $I_n$:
\begin{align}
    Q_s \langle J_1...J_n\rangle&=0 && \Longleftrightarrow && \delta_\xi \trA{[\wv\star ...\star \wv]}=0\,.
\end{align}
Secondly, there can be some global symmetry. For example, in case of $O(M)$ the currents carry nontrivial rank two representations of $O(M)$ such that $J^{ij}_s=(-)^s J^{ji}_s$. The corresponding higher spin algebra is trivially embedded into $\hs \otimes \mathrm{Mat}_M$, where $\hs$ is the higher spin algebra of one complex scalar field (or other fields). Note that the traces enjoy the cyclic symmetry to begin with, i.e. they correspond to correlators in vector models with some $U(M)$ global symmetry. The cases with different or no flavour symmetry are obtained by adding the necessary permutations and by projecting the wave functions $\wv$. Thirdly, there is one implicit parameter, $1/N$, which determines the ratio between connected and disconnected pieces in the correlation functions. It is an external (phenomenological) parameter at the moment and it has nothing to do with higher spin algebra $\hs$. Lastly, the higher spin algebras are usually simple and rigid (= admit no deformations as associative algebras).

\section{Deformed Higher Spin Symmetry}
\label{sec:broken}
Before tackling the problem of deformation of higher spin symmetries, let us abstract the key features of the exact higher spin symmetry. We will only need the fact that there is a Lie algebra $\hsdouble$ together with its action on some module. The Lie algebra originates from an associative algebra $\hsdouble$ with the same name. In practice, the module is just the adjoint module, i.e. $\hsdouble$ itself, and incorporates both the higher spin algebra $\hs$ and bilinear operators $J_s$, see \cite{Sharapov:2018kjz} for more details. Before deformation, the higher spin algebra $\hs$ acts on the higher spin currents $J\sim \hs$ which generate the algebra which acts on the currents, and so on. While in the case of unbroken higher spin symmetries (or in the case of any other usual Lie symmetry) the symmetry algebra can be abstracted and studied independently of various modules it can act on, the deformation of the higher spin symmetry makes the separation of the algebra from its module impossible and they have to deform together. A suitable mathematical concept here is that of a Lie algebroid, which is a special case of $L_\infty$-algebras. It covers the case of a module over a Lie algebra, but goes far beyond that. The reason is that the non-conservation operator on the r.h.s. of $\pl \cdot J=g[JJ]$ is built out of $J_s$ themselves, $g\sim 1/N$. Large-$N$ is important in order to be able to use the classical approximation for composites $[JJ]$. Had the non-conservation operators been just some other operators $O_s$, i.e. not related to $J$'s, the symmetry would have disappeared.

Mathematically, this idea can be cast into the form of an $L_\infty$-algebra, which, in fact, originates from a certain $A_\infty$-algebra by means of anti-symmetrization. The starting point is that we have an associative algebra $\hsdouble$ with product $\star$ and another copy of $\hsdouble$ understood as $\hsdouble$ bi-module. As a graded vector space our $A_\infty$-algebra has two homogeneous subspaces: one in degree $-1$ and the other in degree $0$. We denote them by $V_{-1}$ and $V_0$, respectively. The former accommodates  the algebra, while the latter is reserved for the module. The nonzero structure maps are given by\footnote{We collected the basic definitions in Appendix \ref{app:ALinfinity}. }
\begin{align}
    m_2(a,b)&= a \star b\,, & m_2(a,u)&= a\star u\,, & m_2(u,a)&= -u\star a
\end{align}
for all $a,b\in V_{-1}$ and $u\in V_0$.
The term `algebra together with its bi-module' are now encoded in the Stasheff identity:
\begin{align}
    m_2(m_2(x,y),z)+(-)^{|x|}m_2(x,m_2(y,z))&=0 &&\Longleftrightarrow && \gers{m_2,m_2}=0\,,
\end{align}
where $x,y,z$ are all possible triplets of vectors of $V=V_{-1}\bigoplus V_{0}$. The associated $L_\infty$-algebra has two maps $l_2(a,b)=[a,b]_\star$ and $l_2(a,u)=a\star u-u\star a$. 

As was first observed in \cite{Sharapov:2018kjz}, the initial data above, i.e. an $A_\infty$-algebra with only $m_2$ being nonzero to encode an algebra together with its bi-module, can be deformed whenever the underlying associative algebra admits a deformation as associative algebra. This means that there exists  a one-parameter family of associative algebras $\hs_{\up}$ such that $\hs_{\up=0}\sim \hsdouble$ (which allows us to abuse $\hs_{\up}$). The product in $\hs_{\up}$ can be expanded in powers of the formal deformation parameter $\nu$:
\begin{align}\label{absdeformed}
    a\prd b= a\star b+\sum_{k>0}\phi_k(a,b)\up^k\,.
\end{align}
Here the bilinear operators $\phi_k$ obey certain relations that follow from the associativity of $\prd$: 
\begin{align}\label{assrel}
    \sum_{i+j=n} \phi_i(\phi_j(a,b),c)-\phi_i(a,\phi_j(b,c))=0\,,
\end{align}
where $\phi_0(a,b)\equiv a\star b$. In particular, $\phi_1$ is a nontrivial Hochschild two-cocycle of the algebra $\hsdouble$. With the help of  $\phi_k$'s and explicit formulas from \cite{Sharapov:2018kjz}, one can construct higher structure maps $m_{k>2}$ in such a way that $m=m_2+m_3+...$ obeys the $A_\infty$-relations $\gers{m,m}=0$. For example, for the nontrivial $m_3$'s one finds
\begin{align*}
    m_3(a,b,u)&=\phi_1(a,b)\star u\,,  & m_3(a,u,v)&=\phi_1(a,u)\star v\,, & m_3(u,a,v)&=-\phi_1(u,a)\star v\,,
\end{align*}
where $a,b\in V_{-1}$ and $u,v,w\in V_{0}$. To illustrate a bit more, some of $m_{4}$, $m_5$ read
\begin{align}\label{somemexamples}
    \begin{aligned}
    &\,m_4(a,b,u,v)=\phi_2(a,b)\star u\star  v +\phi_1(\phi_1(a,b),u)\star v\,,\\
    &m_5(a,b,u,v,w)=\phi_1(\phi_1(\phi_1(a,b),u),v)\star w+\phi_2(\phi_1(a,b),u)\star v\star w+\\ &\qquad +\phi_1(\phi_2(a,b),u)\star v\star w +\phi_1(\phi_2(a,b)u,v)\star w+\phi_3(a,b)\star u\star v\star w\,.
\end{aligned}
\end{align}
There is a number of different constructions found in \cite{Sharapov:2018kjz} that lead to explicit description of all $m_n$. As a result, there is an associated $L_\infty$-algebra with maps $l_n$ obtained via anti-symmetrization of $m_n$:
\begin{align}
    l_n(x_1,...,x_n)&= \sum_{\sigma\in S_n} (-)^{\kappa} m_n(x_{\sigma_1},...,x_{\sigma_n})\,,
\end{align}
where $(-1)^\kappa$ is the standard Koszul sign factor deduced from the permutation of $x$'s.

To recapitulate, the main result of \cite{Sharapov:2018kjz} is that one can construct $A_\infty$- and $L_\infty$-algebras that deform the initial data given by an associative algebra $\hs_0$ and its (adjoint) bi-module. The structure maps are built from the deformation $\hs_{\up}$ of $\hs_0$. We are interested in the deformation of the algebra of charges $Q_s$ together with its action on $J_s$. Therefore, we need an $L_\infty$-deformation and an important question whether all such deformations originate this way, i.e. from the deformation of the underlying associative algebra. In general, this does not need to be the case. Deformations of the $L_\infty$-structure are encoded by certain classes of Chevalley--Eilenberg cohomology. Whenever the underlying Lie algebra comes from an associative one, there is a link from the Hochschild to Chevalley--Eilenberg cohomology through the cyclic cohomology, but it is a bit indirect. An important technical assumption in \cite{Sharapov:2018kjz} is that the deformation has to survive once $\hs_0$ gets extended to $\hs_0 \otimes \mathrm{Mat}_M$. In the CFT terms this means that the underlying free CFT can always be extended with some global symmetries,  $U(M)$ or $O(M)$, that effectively replaces $\hs_0$ with $\hs_0\otimes \mathrm{Mat}_M$. 

In the case of free $3d$ scalar and free $3d$ fermion CFT's, where the underlying algebra is the Weyl algebra $A_2$, more detailed answers can be obtained and the matrix extensions can be dropped: it can be shown that the relevant Chevalley--Eilenberg cohomology can be reconstructed from the Hochschild cohomology of the Weyl algebra. This was computed in \cite{Sharapov:2020quq}. In particular, Table 5 of \cite{Sharapov:2020quq} implies that, upon restriction to even spins $J_s$, the $A_\infty/L_\infty$-algebras depend on two parameters, i.e. $\hsdouble$ admits two independent and mutually compatible deformations. Again, the deformed algebra is given by the tensor product $\hs_{\upA,\upB}=\mathcal{A}_{\upA}\otimes \mathcal{A}_{\upB}/\mathbb{Z}_2$. The underlying associative algebra $\mathcal{A}_\up$, of which $\mathcal{A}_0$ is a particular case, has been known for long \cite{Wigner1950, Yang:1951pyq, Boulware1963, Gruber, Mukunda:1980fv} as para-bose oscillators:\footnote{This algebra is also closely related to anyons, see e.g. \cite{Engquist:2008mc}. A relation to anyons was also pointed out recently in \cite{Gandhi:2021gwn} from a completely different perspective. }
\begin{align}
  \mathcal{A}_\up\ni f(q,p,K)&: &  [q, p]&= i(1+\up K)\,, & KqK&=-q\,,& KpK&=-p\,, & K^2&=1\,.
\end{align}
The Hermiticity of the deformation forces $\upA=\upB^*$ and we can put  $\upA=\up e^{i\theta}$, $\upB=\up e^{-i\theta}$ for some real $\up$ and $\theta$. These are to be related to $1/\tilde N$ and another phenomenological parameter, called $\tilde\lambda$ in \cite{Maldacena:2012sf}, $\nu\sim {\tilde N}^{-1}$, $\cos^2 \theta=(1+\tilde\lambda^2)^{-1}$. Therefore, \cite{Sharapov:2020quq} gives a proof that the deformed higher spin symmetry brings in one more parameter on top of $N$. This new parameter is not present in the free limit. In the microscopical realization via Chern--Simons matter theories the parameters are $N$ and level $k$. In the large-$N$ limit one finds $\theta= \tfrac{\pi}2\tfrac{N}{k}$, $\up\sim \tilde{N}^{-1}$, $\tilde{N}=2N\tfrac{\sin \pi \lambda}{\pi \lambda}$. While $N$ and $k$ are real (moreover, they are quantized), it might be interesting to consider the most general deformation with two complex parameters, in particular, where $N/k$ is taken imaginary and $|\theta|$ is large.\footnote{The limit should correspond to Chiral Higher Spin Gravity \cite{Skvortsov:2018uru}, which is a unique perturbatively local theory \cite{Metsaev:1991mt,Metsaev:1991nb,Ponomarev:2016lrm} and it is at least one-loop finite \cite{Skvortsov:2020wtf,Skvortsov:2020gpn}.\label{ft:chiral}}

Let us also formulate the invariants \eqref{freeTr} of the deformed higher spin symmetry in the language of $L_\infty$-algebra. Consider a complex-valued function $F$ defined on the even subspace $V_0\subset V$ by a  series 
\begin{equation}\label{F}
    F=F_1(u)+\frac12F_2(u,u)+\frac13F_3(u,u,u)+\ldots\,.
\end{equation}
We say that the function $F$ is $l$-invariant if 
\begin{equation}\label{dF}
    (\delta_a F)(u)\equiv\sum_{k=2}^\infty\sum_{n=1}^\infty F_n(l_k(a, u,\ldots,u),u,\ldots,u)=0
\end{equation}
for all $a\in V_{-1}$. For graded Lie algebras ($l_k=0, \forall k>2$), 
the l.h.s. of this equation reproduces the standard action of the Chevalley--Eilenberg differential on  
scalar cochains in degree $0$. Formula (\ref{dF}) allows one to identify  the $l$-invariant functions on $V_0$ with certain classes of $L_\infty$-cohomology. In physical terms, it is a functional of higher spin currents that is invariant under the deformed higher spin symmetry. 

There is a close relationship between the $l$-invariant functions (\ref{F}) and invariants of the underlying $A_\infty$-algebra. As is explained  in Appendix \ref{app:ALinfinity}, see Eq. (\ref{A6}), the operator $L_m$ defines an endomorphism on the space of 
multilinear functions on $V=V_{-1}\oplus V_0$. 
A function $S$ is said to be $m$-invariant if $L_mS=0$. We are interested in $m$-invariant functions that obey a couple of additional conditions. First, each homogeneous component in the expansion
\begin{equation}\label{SS}
 S=S_1(x_1)+S_2(x_1,x_2)+S_3(x_1,x_2,x_3)+\cdots  
\end{equation}
is supposed to satisfy the cyclicity condition (\ref{A8}).
Second, we require that the functions $S_n$ vanish whenever at least one  of their arguments belongs to $V_{-1}$. Hence, $S_n$ is a cyclic invariant function entirely supported on the even subspace $V_0$, i.e. on higher spin currents.  We claim that every such function (\ref{SS}) gives rise to an $l$-invariant function (\ref{F}) by setting
\begin{equation}
    F_n=S_n(u,\ldots,u)\,.
\end{equation}
In the next section and  Appendix \ref{app:proof}, we explain how to construct a series of $l$-invariants of the form
\begin{equation}\label{ln}
    I_n(u)=\frac1n\mathrm{Tr}_0(\underbrace{u\star\cdots \star u}_{n})+O(u^{n+1})\,,
\end{equation}
whenever the corresponding $L_\infty$-algebra comes from a one-parameter family of associative algebras with trace. Clearly, the leading term in (\ref{ln}) defines a Chevalley--Eilenberg cocycle for the $\star$-commutator Lie algebra.

\section{Correlation Functions}
\label{sec:correlators}
As it was already explained in Section \ref{sec:unbroken}, correlation functions in a free CFT are given by invariants $I_n=\tfrac1n\trA_\star[{\wv\star...\star \wv}]$ of the higher spin symmetry. As the higher spin symmetry gets deformed, $I_n$'s cease to be invariant and need modifications. A priori it is not even clear if the invariants survive the deformation. 

Two important facts follow from the classification of higher spin invariants and obstructions to deformations thereof obtained in \cite{Sharapov:2020quq}, see Table 1. Firstly, there is exactly one invariant $I_n$ of type $\wv^n$ and it starts as $\Trace{\star}{\wv\star...\star \wv}$. Secondly, the deformation is unobstructed since the cohomology group containing obstructions to deformation of $I_n$ is empty. Therefore, we can conclude that each $I_n$ can be deformed to remain invariant under the slightly-broken higher spin symmetry. Moreover, the deformation is unique -- there are no coboundaries that can be added to $I_n$. The last property, however trivial its derivation, is of paramount significance for CFT: correlation functions are completely fixed by higher spin symmetry and have no other hidden free parameters. This implies the three-dimensional bosonization duality in the following sense: whatever the microscopical realization of CFT models is, if the higher spin symmetry is deformed/broken in the way we described, there is a unique answer for the correlation functions (up to two parameters for the models with higher spin currents of even spins, for simplicity, which are to be relate to $N$ and level $k$ in the microscopical realization).

However, as usual such statements do not come with an explicit form of the deformed invariants. More generally, it would be interesting to construct the deformed invariants, i.e. exact invariants of the slightly-broken higher spin symmetry, with some minimal assumptions on the deformation of the underlying associative algebra. We know that the entire $L_\infty$-algebra that describes the deformed higher spin symmetry results from a deformed associative algebra, which equips us with $\phi_n$, \eqref{absdeformed}. Since the free CFT correlators result from the trace $\Trace{\star}{\bullet}$ let us assume that the deformed higher spin algebra also has a trace 
\begin{equation}\Trace{\prd}{\bullet}=
\Trace{0}{\bullet}+\nu\, \Trace{1}{\bullet}+\nu^2\,\Trace{2}{\bullet}+\ldots
\end{equation}
which reduces to $\Trace{0}{\bullet}\equiv \Trace{\star}{\bullet}$ for $\nu=0$. By definition, the trace obeys 
\begin{align}\label{tracedef}
    \Trace{\prd}{a\prd b-b\prd a}&=0 && \Longleftrightarrow&&  \sum_{i=0}^{i=n} \Trace{n-i}{\phi_{i}(a,b)-\phi_{i}(b,a) }=0\,.
\end{align}
That the deformed algebra admits a trace is certainly true for the case of $3d$ vector models, but it is also true for all higher spin algebras we are aware of. Therefore, the assumption of $\hs_{\up}$ having a trace $\Trace{\prd}{\bullet}$ seems reasonable.

Below, we explicitly construct invariants of the deformed higher spin symmetry under the only assumption that the deformed algebra admits a trace, which is, in particular, true for the case of Chern--Simons matter models. We first consider examples of such deformations to clarify the general structure of the all-order result. Let us start with the initial $n$-point function\footnote{As before, we consider cyclic invariant expressions, which corresponds to $U(M)$ global symmetry, see also the end of section \ref{sec:unbroken}. To get colorless expressions one simply needs to add permutations. }
\begin{align}
    I_n&=\tfrac1{n}\Trace{0}{\wv^{\star (n)}}\equiv \tfrac1{n}\Trace{0}{\wv\star \wv\star ...\star \wv}
\end{align}
Let us recall that it is invariant up to the leading order due to $\delta_0 \wv=[\xi,\wv]$ and $\Trace{0}{[a,b]_\star}\equiv0$. At the next order we find 
\begin{align}\label{noninvA}
    \delta_1 \Trace{0}{\wv\star \wv...\star \wv}&= \sum_i \Trace{0}{\wv\star ... \star l_3(\xi,\wv,\wv)\star ...\star \wv}\,.
\end{align}
Now it can be seen that the non-invariance \eqref{noninvA} can be compensated  by two types of terms:
\begin{align}
    \tfrac1{n+1}\Trace{1}{\wv^{\star (n+1)}}+\sum_{i=0}^{i=n-1}\tfrac1{n+1}\Trace{0}{\wv^{\star (i)}\star \phi_1(\wv,\wv^{\star (n-i)})}\,.
\end{align}
The process can be continued and it is easy to see that knowing $\phi_n$ and $\Trace{n}{\bullet}$ is sufficient to find an explicit form. It is convenient to define $\wv^{\prd (n)}_k$ as the coefficient of $\nu^k$ in the expansion
\begin{align}\label{expansionJ}
    \wv^{\prd (n)}&\equiv \wv\prd ...\prd \wv=\wv^{\star (n)} + \up\sum_{i=0}^{i=n-1}\wv^{\star (i)}\star \phi_1(\wv,\wv^{\star (n-i)})+ \mathcal{O}(\up^2) \,.
\end{align}
It should be noted that due to the identities among $\phi_k$, which result from the associativity \eqref{assrel}, there can be many ways to write the same expression. In particular, the second group of terms in \eqref{expansionJ} can be rewritten in many equivalent forms. For example, grouping all brackets on the right, i.e. $(x\prd ( ...  (c \prd (a\prd b)))   )$ we find
\besubeqs
\begin{align}
    \wv^{\prd 2}_1&=\phi_1(\wv,\wv)\,,\\
    \wv^{\prd 3}_1&=\phi_1(\wv,\wv\star \wv) +\wv\star \phi_1(\wv,\wv)\,,\\
    \wv^{\prd 3}_2&=\phi_2(\wv,\wv\star \wv) +\phi_1(\wv, \phi_1(\wv,\wv))+\wv\star\phi_2(\wv, \wv)  \,. 
\end{align}
\esubeqs
With a kind word and brute force we find the following expressions as prolongations of the lowest higher spin symmetry invariants to the first few orders: 
\begin{align*}
    I_1[\wv]&=\Trace{0}{\wv}+\tfrac12\Trace{0}{\wv^{\prd2 }_1}+\tfrac12\Trace{1}{\wv^{\prd 2}_0}+\tfrac13\Trace{0}{\wv^{\prd 3 }_2}+\tfrac13\Trace{1}{\wv^{\prd 3 }_1}+\tfrac13\Trace{2}{\wv^{\prd 3 }_0}+...\,,\\
    I_2[\wv]&=\tfrac12\Trace{0}{\wv\star \wv}+\tfrac13\Trace{0}{\wv^{\prd3 }_1}+\tfrac13\Trace{1}{\wv^{\prd 3}_0}+\tfrac14\Trace{0}{\wv^{\prd 4}_2}+\tfrac14\Trace{1}{\wv^{\prd 4}_1}+\tfrac14\Trace{2}{\wv^{\prd 4}_0}+...\,,\\
    I_3[\wv]&=\tfrac13\Trace{0}{\wv\star \wv\star \wv}+\tfrac14\Trace{0}{\wv^{\prd 4}_1}+\tfrac14\Trace{1}{\wv^{\prd 4}_0}+...\,,\\
    I_4[\wv]&=\tfrac14\Trace{0}{\wv\star \wv\star \wv\star \wv}+\tfrac15\Trace{0}{\wv^{\prd 5}_1}+\tfrac15\Trace{1}{\wv^{\prd 5}_0}+...\,.
\end{align*}

\paragraph{From $I_1$ to $I_n$.} It turns out that all information about higher order invariants is already present in the deformation of the first one, $I_1=\Trace{0}{\wv}$+.... Let us peel off the trace and organize various terms that appear in the deformed $I_1$ in a table:
\begin{align}
    & \wv && \tfrac12 \wv^{\prd 2}_0\equiv\tfrac12 \wv\star \wv && \tfrac13 \wv^{\prd 3}_0 && \tfrac14 \wv^{\prd 4}_0 &&  ... && \tfrac{1}{n+1}\wv^{\prd (n+1)}_0 \notag\\ 
    & 0 && \tfrac12 \wv^{\prd 2}_0\equiv\tfrac12\phi_1(\wv,\wv) && \tfrac13\wv^{\prd 3}_1 && \tfrac14\wv^{\prd 4}_1 && ...&& \tfrac{1}{n+1}\wv^{\prd (n+1)}_1\notag\\
    & 0 && 0 && \tfrac13\wv^{\prd 3}_2 && \tfrac14\wv^{\prd 4}_1 && ... && ...\label{tableInv}\\
    & 0 && 0 && 0 && ... && ... && ...\notag\\
    & 0 && 0 && 0 && ... && ... && \tfrac{1}{n+1}\wv^{\prd (n+1)}_n\notag
\end{align}
where we have already displayed the right terms, but this is not important for the argument below, i.e. it will work irrespective of the concrete form of the terms in the table encoding $I_1$. In the table going right increases the number of $\wv$ and going downwards increases the total degree of $\phi$'s.\footnote{To be precise, let $|\phi_k|=k$, so that $\phi_1(\wv,\phi_1(\wv,\wv))$ and $\phi_2(\wv,\wv\star \wv)$ both have degree two.} In order to construct deformed $I_1$ one applies $\Trace{0}{\bullet}$ to the leftmost diagonal adds $\Trace{1}{\bullet}$ of the second diagonal, etc. 

An interesting property is that one can construct all $I_n$ once we know $I_1$. To do so one needs to erase $k$ diagonals from the left, then contract the first diagonal left with $\Trace{0}{\bullet}$, the second one with $\Trace{1}{\bullet}$, etc. and add everything up. Note that due to the shift of the index $k$ of trace $\Trace{k}{\bullet}$ the resulting expression is not equal to $I_1$ with some terms removed. The resulting expression is invariant under the deformed higher spin symmetry! Indeed, the invariance of $I_1$ implies that $\delta_\xi I_1$ has such a form that various parts combine into the trace identities \eqref{tracedef}, i.e. what is inside $\Trace{i}{\bullet}$ forms different parts of the deformed commutator $[a,b]_\prd=a\prd b-b\prd a$. By counting various orders of $\wv$ and $\phi$ it is easy to see that the recipe above does work. 

Since in the case of $3d$ vector models there is a unique invariant $I_n$, the procedure above generates exactly $I_n$ from $I_1$. In particular, following the recipe we get $I_n$ that starts with $\tfrac1{n}\Trace{\star}{\wv^{\star(n)}}$, as required. To summarize, we just need to find $I_1$ in order to derive all $I_n$. 

\paragraph{Generating function. }
The free CFT correlators' generating function is 
\begin{align}
    W[\wv]&=\Trace{\star}{\log_\star [1-\nu^{-1}\wv]}\,,
\end{align}
where we introduced $\nu$ by hand to count orders for the ease of comparing with the deformed invariants below, cf. footnote \ref{normalization}. The first few orders of the deformed invariants suggest that the complete answer reads
\begin{align}\label{completecor}
    I_n[\wv]&= \sum_{k=0} \frac{1}{n+k}\sum_{i=0}^{i=k}\Trace{k-i}{\wv^{\star (n+k)}_i}\,.
\end{align}
With some abuse of notation the latter formula can be compactly written as
\begin{align}
    W_{\text{SBHS}}[\wv]&=\sum_n I_n[\wv]=\Trace{\prd}{\log_\prd [1-\up^{-1} \wv]} \, \mod \, \text{irrelevant}\,,
\end{align}
where modding out irrelevant terms is very important. By irrelevant terms we mean those that do not contribute according to \eqref{completecor}. These are the terms that correspond to zeros in \eqref{tableInv}, which otherwise could have been filled in with $\wv^{\prd (n)}_k$, $k\geq n$. They do not show up for a simple reason: the order of $\phi$ is related to the order in $\wv$. We can also rewrite it as
\begin{align}
    W_{\text{SBHS}}[\wv]&=\sum_n I_n[\wv]=\Trace{\prd}{\log_\prd [1-\up^{-1}\wv]}\big|_{\text{p.p.}}\,,
\end{align}
where both $\Trace{\prd}{\bullet}$ and $\prd$-product are expanded in formal deformation parameter $\up$ and p.p. means the principal part of the formal Laurent series in $\up$. The proof of this formula is heavily based on the tools of non-commutative geometry and we defer part of it to Appendix \ref{app:proof}. 

We note that wherever there is a one-parameter $\hs_{\up}$ family of algebras together with the corresponding $\Trace{\nu}{\bullet}$, the parameter enters as an overall factor only. Therefore, it corresponds, in some sense, to the effective number of degrees of freedom $\tilde N$. If there are several parameters $\up_i$, then the correlators have simple polynomial dependence on $\up_i$. In the case of $3d$ vector models, where there is a two-parameter family of deformations, provided the Hermiticity conditions are imposed, the absolute value $|\nu|$ corresponds to $\tilde{N}$, while the second, phase-like, parameter enters the structure of the correlators. As was already noted, it might be interesting to formally extend to the complex domain, which brings more real parameters to play with.

\paragraph{First few correlators.} Now let us collect terms of the same order in $\wv$. They should give independent contributions to the corresponding correlation functions. For example, 
\begin{align*}
    1&: && \Trace{0}{\wv}\,,\\
    2&: && \tfrac12\Trace{0}{\wv\star \wv}+\tfrac12\Trace{0}{\wv^{\prd2 }_1}+\tfrac12\Trace{1}{\wv^{\prd 2}_0}\equiv \tfrac12\Trace{0}{\wv\star \wv}+\tfrac12\Trace{0}{\phi_1(\wv,\wv)}+\tfrac12\Trace{1}{\wv\star \wv}\,,\\
    3&: &&\tfrac13 \Trace{0}{\wv\star \wv\star \wv}+\tfrac13\Trace{0}{\wv^{\prd3 }_1}+\tfrac13\Trace{1}{\wv^{\prd 3}_0} +\tfrac13\Trace{0}{\wv^{\prd 3 }_2}+\tfrac13\Trace{1}{\wv^{\prd 3 }_1}+\tfrac13\Trace{2}{\wv^{\prd 3 }_0}
     \,,\\
    4&: &&\tfrac14 \Trace{0}{\wv\star \wv\star \wv\star \wv}+\tfrac14\Trace{0}{\wv^{\prd 4}_1}+\tfrac14\Trace{1}{\wv^{\prd 4}_0} +\tfrac14\Trace{0}{\wv^{\prd 4}_2}+\tfrac14\Trace{1}{\wv^{\prd 4}_1}+\tfrac14\Trace{2}{\wv^{\prd 4}_0}+\\
    &&&\qquad \qquad+\tfrac14\Trace{0}{\wv^{\prd 4}_3}+\tfrac14\Trace{1}{\wv^{\prd 4}_2}+\tfrac14\Trace{2}{\wv^{\prd 4}_1}+\tfrac14\Trace{3}{\wv^{\prd 4}_0}\,.
\end{align*}
It is hard to say what different terms are responsible for without actually computing them, save for the first ones representing free CFT's correlators. Nevertheless, let us speculate a little bit. First of all, the deformed higher spin symmetry leads to a finite number of terms added to each free $n$-point correlation function. Some of these terms can vanish or be proportional to each other for specific $\wv$'s. One general remark is that the invariants should be fed with appropriate wave functions $\wv$ that, among other things, contain information about the topology of spacetime, e.g. whether it is flat or it has some thermal circle. For example, the one-point function has to vanish in flat space, but it does have in a thermal background. It is also clear that the correlation functions depend analytically on the phenomenological coupling constants and there cannot be any functions of cross-ratios with nontrivial dependence on these parameters. 

\section{Conclusions and Discussion}
\label{sec:conclusions}
The main result of the paper is remarkably simple and calls for a better explanation: slightly-broken higher spin symmetry admits a family of invariants $W_{\text{SBHS}}[\wv]$ that can compactly be written as
\begin{align}
    W_{\text{SBHS}}[\wv]&=\Trace{\prd}{\log_\prd [1-\up^{-1}\wv]}\big|_{\text{p.p.}}\,, & W_\text{free}[\wv]&=\Trace{\star}{\log_\star [1-\up^{-1}\wv]}\big|_{\text{p.p.}}\,,
\end{align}
where p.p. means the principal part of the formal Laurent series and the deformation parameter $\up$ is a gadget to count the order of invariants. $W_{\text{SBHS}}$ is a deformation of the free CFT's correlators $W_\text{free}$. In $W_\text{free}$ the dependence on $\up$ is displayed explicitly, while in $W_{\text{SBHS}}$ it is also present in the trace and $\prd$-product. When written this way $W_{\text{SBHS}}$ is very similar to the result $W_\text{free}$ in the corresponding free CFT. The final result, perhaps, implies that all CFT's with a slightly-broken higher spin symmetry are quasi-free in the large-$N$ limit, which is supported by the simplicity of the concrete results in Chern--Simons vector models collected so far, e.g. \cite{Giombi:2011kc, Maldacena:2012sf,Aharony:2012nh, GurAri:2012is,Giombi:2016zwa,Li:2019twz,Kalloor:2019xjb,Turiaci:2018nua,Jain:2021gwa,Silva:2021ece} for an incomplete list of results. 

A clear omission of the present paper is the lack of concrete examples where the correlation functions are computed down to the final expressions in terms of spacetime points and polarization vectors, save for the free CFT's correlators in \cite{Colombo:2012jx,Didenko:2012tv,Didenko:2013bj, Bonezzi:2017vha}. We  hope this is compensated by the generality of the main statements: the results apply to every physical model, not necessarily a CFT, where the action of a symmetry gets deformed together with its action in a certain module, so that the resulting structure is $L_\infty$. The main technical assumption, which is certainly true in the CFT context, is that the initial (Lie) symmetry algebra originates from an associative one. 

In the general case, the deformation of a symmetry together with its module to a full-fledged $A_\infty/L_\infty$-algebra is governed by the Chevalley--Eilenberg cohomology, which, to a large extent, can be be reduced to a much simpler Hochschild cohomology \cite{Sharapov:2020quq}. The dimension of the corresponding cohomology group is equal to the number of independent parameters on the CFT side, provided there are no obstructions. One parameter always plays the role of $1/N$ and enters as an overall factor that counts the order of deformation. The rest of the parameters can enter in a more interesting way. Provided the number of parameters is finite, the correlation functions get a finite number of corrections due to the deformed higher spin symmetry. This implies that the $n$-point correlators have a very specific form of a sum of several structures multiplied by the free parameters, i.e. one cannot have a function of cross-ratios that depends nontrivially on the coupling constants.

As we already mentioned, for $3d$ vector models all the relevant cohomology was computed in \cite{Sharapov:2020quq}. Taking the simplest spectrum of higher spin currents with only even spins, Table 1 there implies that there is exactly one invariant of the higher spin symmetry that starts as $\Trace{0}{\wv\star...\star \wv}$ and it is unobstructed. Table 5 implies that the deformation to $L_\infty$-algebra depends on two (phenomenological) parameters. Therefore, in any $3d$ CFT with such a spectrum all correlation functions should be given by $W_{\text{SBHS}}$ provided an appropriate identification between the microscopical parameters and the phenomenological ones is made. This implies the three-dimensional bosonization duality. However, a proof of the concept requires one to compute at least some of the correlation functions, which should also explain what different terms correspond to.

The approach advocated in this paper is based entirely on symmetry. The $L_\infty$-algebra contains full information that can be extracted from the non-conservation equation $\pl \cdot J=g[JJ]$. However, the double-trace operators $[JJ]$ get renormalized in a certain way away from the large-$N$ limit and the `semi-classical' large-$N$ approach should not be valid anymore. In particular, $W_{\text{SBHS}}$ should not contain any anomalous dimensions. Nevertheless, the LO anomalous dimensions of higher spin currents $J_s$ can be extracted from the non-conservation equation \cite{Anselmi:1998ms,Skvortsov:2015pea,Giombi:2016hkj,Giombi:2016zwa,Giombi:2017rhm}.\footnote{The NLO anomalous dimensions in the critical vector and Gross-Neveu models are known \cite{Manashov:2016uam,Manashov:2017xtt}.} It would be important to understand if and how the approach can be extended beyond large-$N$.\footnote{Anomalous dimensions of higher spin currents in the Ising model, $N=1$, are also quite small. Moreover, they get smaller and smaller as the spin grows \cite{Komargodski:2012ek}.}

Another aspect of the approach that calls for a better understanding are wave-functions $\wv$, which are well-known for the $3d$ case \cite{Colombo:2012jx,Didenko:2012tv,Didenko:2013bj, Bonezzi:2017vha}, see also \cite{Iazeolla:2008ix,Didenko:2012vh} for some results on any $d$. They have a number of remarkable properties, e.g. they are extremal projectors. Fundamentally, $\wv$ represents states of the form $|\phi\rangle\langle\phi|$, i.e. it gives an embedding of the tensor product state $|\phi\rangle\otimes |\phi\rangle$ inside the higher spin algebra together with other primary operators $J_s$. 

An important related question left aside is fixing the diffeomorphism symmetry $\wv\rightarrow \wv+f(\wv)$, which is inherent in the $L_\infty$-algebras (or  associated {\it $Q$-manifold}, a term that goes better with diffeomorphisms) by definition, but not in CFT. Wave-function $\wv$ has to transform canonically under the conformal symmetries, i.e. as a quasi-primary operator, which is a property of $\wv$ and is not built into $L_\infty$-algebras. By construction, the module on which the higher spin algebra acts contains quasi-primary operators together with descendants. However, one would like to compute correlation functions of the quasi-primary operators $J_s$ rather than descendants. Therefore, we need to find a distinguished coordinate system on the $Q$-manifold that makes the action of the conformal symmetry on $\wv$ canonical. This should not pose any conceptual difficulties since for any $J_s$ there is always a finite number of descendants of $J_{s'<s}$ that can mix with it.  

An interesting application of the present results is to the higher spin gravity duals of the vector models \cite{Sezgin:2002rt,Klebanov:2002ja,Sezgin:2003pt,Leigh:2003gk}, which exhibit certain non-locality \cite{Bekaert:2015tva,Sleight:2017pcz,Ponomarev:2017qab} that prevents one from extracting correlation functions \cite{Giombi:2009wh,Giombi:2010vg,Boulanger:2015ova,Skvortsov:2015lja} except at the lowest order. These non-localities preclude any bulk intrinsic definition of higher spin gravity duals of vector models within the field theory approach at the moment.\footnote{Given an $L_\infty$-algebra one can always write down a sigma-model, which leads to formally consistent classical equations \cite{Vasiliev:1988sa,Vasiliev:1990cm} that have some features of higher spin gravity. Already at the cubic order there are some issues \cite{Giombi:2009wh,Giombi:2010vg,Boulanger:2015ova,Skvortsov:2015lja}. Had the corresponding higher spin gravity existed in the sense of having a well-defined prescription to compute holographic correlation functions, it would have been possible to give a constructive proof of the bosonization duality by proving the uniqueness of the theory and by computing the correlation functions. Two exceptions should be mentioned. (1) the Collective Dipole approach \cite{deMelloKoch:2018ivk,Aharony:2020omh} reconstructs a theory in $AdS$ with some properties of the sought for higher spin gravity. It also gives a well-defined prescription to compute correlation functions. However, its extension to the duals of Chern--Simons vector models is not yet available. (2) There is a certain non-unitary limit of Chern--Simons vector models that has a well-defined local higher spin gravity dual, cf. footnote \ref {ft:chiral}. The non-existence of many of the most interesting higher spin gravities is not an obstacle for exploring the higher spin symmetry directly on the boundary, where everything is well-defined. } The idea of applying higher spin symmetry on the CFT side gives a concrete principle of how to calibrate the $L_\infty$ structure maps, which can later be used to define the higher spin gravity dual. This, however, cannot completely resolve the non-locality issues (the theory still has to be more non-local than the field theory approach allows for \cite{Bekaert:2015tva,Sleight:2017pcz,Ponomarev:2017qab}) and a further prescription on how to compute correlation functions should be needed. 

A drastic difference between CFT and AdS applications of higher spin symmetry can be explained as follows. Starting from the quartic order the most general AdS interaction vertex is an infinite series in derivatives. Likewise, on the CFT side a function of cross-ratios can accommodate infinitely many expansion coefficients. The first bulk problem is that gauged higher spin symmetry, i.e. duality to vector models, requires non-local interactions. These interactions are incomplete by themselves and call for a prescription of how to deal with them \cite{Bekaert:2015tva}. The second bulk problem is the possibility to perform field redefinitions, which now can be non-local as well and can affect the result. None of these problems is present on the CFT side. Just the dilation symmetry fixes the number of derivatives and makes the action of symmetry $[Q,J]=J+...$ well-defined on quasi-primary operators $J_s$.    
From this point of view, the two sides of the AdS/CFT correspondence are not equal: the CFT correlation functions are much less sensitive to a particular form of $L_\infty$-algebra than the vertices of bulk theory.

Another application to higher spin gravity is that the correlation functions, being invariants of (deformed) higher spin symmetry \cite{Sharapov:2020quq}, have to be given by the same expressions.\footnote{Historically, the idea of using trace-type invariants was first proposed in \cite{Sezgin:2005pv} to distinguish between different solutions, see also \cite{Engquist:2005yt,Colombo:2010fu,Sezgin:2011hq,Colombo:2012jx}. Note, however, that these invariants are defined not on the higher spin algebra, but on a much bigger algebra, so the relation to the results of the present paper is yet to be clarified, see also \cite{Vasiliev:2015mka} for a related approach.  } It would be important to find a direct proof that e.g. the AKSZ action of \cite{Sharapov:2021drr} leads to the same correlation functions. The invariants of the higher spin algebra are also very close to the vertices of the Collective Dipole action, see e.g. \cite{deMelloKoch:2018ivk}; the relation between the two is discussed in \cite{Neiman:2017mel,Neiman:2018ufb}. We hope that the results of this paper pave a way to extending the Collective Dipole action to Chern--Simons vector models without having to directly integrate the Chern--Simons gauge field out. 

A different idea in the same direction \cite{Skvortsov:2018uru} is to start from Chiral Higher Spin Gravity where the second parameter results from EM-duality transformations, which immediately proves that there is a one-parameter family of theories connecting the parity preserving vector models pairwise (i.e. there is a one-parameter family of CFT's connecting the free fermion CFT to critical vector model and the free boson CFT to Gross-Neveu model), but the uniqueness of the deformation is not obvious at higher orders. In particular, all three-point functions of higher spin currents were found in \cite{Skvortsov:2018uru} and the uniqueness of the solution can be shown at this level. This result supplements \cite{Maldacena:2012sf} with concrete expressions for all three-point functions and gives a simple and independent proof of there being one additional parameter that enters the correlators. Extension to higher orders is simple modulo the proof of uniqueness. 

\section*{Acknowledgments}
\label{sec:Aknowledgements}

We are grateful to Yasha Neiman, Dmitry Ponomarev and Alexander Zhiboedov for useful discussions. The work of A. Sh. was supported by the Ministry of Science and Higher Education of the Russian Federation, Project No. 0721-2020-0033, and by the Foundation for the Advancement of Theoretical Physics and Mathematics “BASIS”. The work of E.S. and P.G. was supported by the Russian Science Foundation grant 18-72-10123 in association with the Lebedev Physical Institute. 

\appendix

\section{Strong Homotopy Algebras}\label{app:ALinfinity}
This section is not aimed at a comprehensive introduction into the theory of strong homotopy algebras, see e.g. \cite{Kajiura:2003ax,Lada:1992wc,Cuntz,stasheff2018linfty}  for reviews, and we just collect the basic definitions. 

Let $V=\bigoplus_{k\in \mathbb{Z}} V^k$ be a $\mathbb{Z}$-graded vector space over $\mathbb{C}$. Consider the space $Hom(TV,V)$ of all maps from the tensor algebra $TV=\bigoplus_n T^nV$ of $V$ to the space $V$ itself. The element of $Hom(T^nV,V)$, called {\it $n$-cochains}, are multilinear functions $f(x_1,x_2,\ldots,x_n)$ on $V$. The $\mathbb{Z}$-grading on $V$ induces that on $Hom(TV,V)$; by definition, 
$$
|f|=|f(x_1,x_2,\ldots,x_n)|-\sum_{k=1}^n|x_k|\,.
$$
The  $\circ$-product of an $n$-cochain $f$ and an $m$-cochain $g$ is a natural operation that nests one map into the other with the usual Koszul signs
\begin{align}\label{comp-prod}
    \begin{aligned}
    (f\circ g)&(x_1\otimes x_2\otimes\cdots\otimes x_{m+n-1} )=\\
     &=\sum_{i=0}^{n-1}(-1)^{|g|\sum_{j=1}^i|x_j|} f(x_1\otimes \cdots\otimes x_i\otimes g(x_{i+1}\otimes \cdots\otimes x_{i+m})\otimes \cdots \otimes x_{m+n-1})\,.
    \end{aligned}
\end{align}
It should be noted that the $\circ$-product is non-associative. Nevertheless, the following bracket, called {\it Gerstenhaber bracket},
\begin{align}\label{G}
   \gers{f,g}=f\circ g-(-1)^{|f||g|}g\circ f\,,
\end{align}
is graded skew-symmetric and obeys the Jacobi identity: 
\begin{align}
   \gers{f,g}&=-(-1)^{|f||g|}\gers{g,f} \,, &&\gers{\gers{f,g},h}=\gers{f,\gers{g,h}}-(-1)^{|f||g|}\gers{g,\gers{f,h}}\,.
\end{align}
Given a $\mathbb{Z}$-graded space $V$ and a sum $m=m_1+m_2+\cdots $ of degree-one maps $m_n: T^nV\rightarrow V$, the $A_\infty$-structure is defined simply as a solution to the Maurer--Cartan equation:
\begin{align}\label{MC}
    \gers{m,m}&=0\,.
\end{align}
Upon expansion  $m=m_1+m_2+\cdots $ the first few relations have a simple interpretation: $m_1$ is a differential, $m_1m_1=0$; $m_2$ is a bi-linear product differentiated by $m_1$ by the graded Leibnitz rule  
\begin{align}
-m_1 m_2(a,b)=m_2(m_1(a),b)+(-1)^{|a|}m_2(a,m_1(b))\,.    
\end{align}
However, $m_2$ is not associative in general, associativity is true up to a coboundary controlled by $m_3$:
\begin{align*}
m_2(m_2(a,b),c)+&(-1)^{|a|}m_2(a,m_2(b,c))+m_1 m_3(a,b,c)+m_3(m_1(a),b,c)+\\
&\qquad+(-1)^{|a|}m_3(a,m_1(b),c)+(-1)^{|a|+|b|}m_3(a,b,m_1(c))=0\,.
\end{align*}

The graded Lie algebra of cochains (\ref{G}) admits a natural representation in the space $Hom(TV,\mathbb{C})$ of multilinear complex-valued functions on $V$. This is defined by the assignment  $f \mapsto L_f$, where 
\begin{equation}\label{A6}
    (L_f S)(x_1,\ldots, x_{q+p-1})=
\end{equation}
$$
\sum_{j=1}^{q}(-1)^{\sum_{i=1}^k |f|(|x_1|+\cdots +|x_{j}|)} S(x_1,\ldots, x_j, f(x_{j+1},\ldots, x_{j+p}),\ldots,x_{q+p-1})
$$
$$
+\sum_{j=q-1}^{p+q-1}(-1)^{(|x_1|+\cdots+|x_j|)(|x_{j+1}|+\cdots +|x_{p+q-1}|)}S(f(x_{j+1},\ldots,x_1, \ldots, x_{j+p-n-1}),\ldots, x_j) 
$$
for all $f\in Hom(T^pV,V)$ and $S\in Hom(T^qV,\mathbb{C})$.  It is not hard to check that 
\begin{equation}
    [L_f,L_g]=L_{\gers{f,g}}\qquad \forall f,g \in Hom(TV,V)\,.
\end{equation}
 This representation is reducible as $Hom(TV,\mathbb{C})$ contains an invariant subspace spanned by multilinear functions obeying the cyclicity condition
\begin{equation}\label{A8}
    S(x_0,\ldots, x_p)=(-1)^{x_p(x_0+\cdots+x_{p-1})}S(x_p,x_0,\ldots,x_{p-1})\,. 
\end{equation}
For more details on the last definitions, see e.g. \cite{Getzler93}.

\section{Sketch of the Proof}\label{app:proof}
The proof consists of two parts: hard and easy.  The hard part relies heavily on general ideas of non-commutative geometry \cite{Cuntz} and cannot be reproduced here in any detail; we will present it elsewhere. In a few words, one can define a flow on the space of $A_\infty$-cohomology that is compatible with a flow for the $A_\infty$-algebra we are dealing with. The final result -- the flow equation -- is easy to write down explicitly and we do this below. It turns out that the $A_\infty$-algebra of Section \ref{sec:broken} is the $\up\rightarrow 0$ limit of a one-parameter family of auxiliary $A_\infty$-algebras \cite{Sharapov:2018kjz}. Taylor coefficients $\phi_k(\bullet,\bullet)$, cf. around \eqref{somemexamples}, can be replaced by derivatives of the deformed product 
$\mu(a,b)=a \prd b=a\star b+ \mathcal{O}(\up)$, \eqref{absdeformed}, according to $\phi_k\rightarrow (k!)^{-1}(\pl_\up)^k\mu$. This operation respects the Maurer--Cartan equation \eqref{MC}. Let us denote the structure maps of this auxiliary $\up$-dependent family of $A_\infty$-algebras as $\boldsymbol{m}_n$. We set
\begin{align}
    \mathfrak{m}_n(v_1,...,v_{n})&\equiv \boldsymbol{m}_n(\pl(v_1),v_2,...,v_{n}) \,,
\end{align}
where $v_i\in A_0$ and $\pl: A_{0}\rightarrow A_{-1}$ is a somewhat trivial map that send elements from $A_{0}$ to $A_{-1}$ by merely changing their degree assignment (note that $A_{0}\sim A_{-1}$ as vector spaces). A useful observation \cite{Sharapov:2018kjz} is that $\boldsymbol{m}_n$ or $\mathfrak{m}_n$ obey the first-order differential  equation 
\begin{align}\label{floweq}
    \dot {\mathfrak{m}}(v,...,v)&= \mathfrak{m}(\mathfrak{m}'(v,...,v),v,...,v)\,,
\end{align}
which reconstructs $\mathfrak{m}(t)=\mathfrak{m}^{(0)}+t\mathfrak{m}^{(1)}+...$ order by order in $t$ such that $\gers{\mathfrak{m}(t),\mathfrak{m}(t)}=0$ and $\mathfrak{m}_2+\mathfrak{m}_3+...=\mathfrak{m}(t=1)$. Hereinafter $f'$ is a short-hand for $\pl_\up f$. The initial condition for (\ref{floweq}) is $\mathfrak{m}_2(v,v)\equiv \mathfrak{m}^{(0)}(v,v)=\mu(v, v)$. For example, solving Eq. (\ref{floweq}) by iterations, one finds
\begin{align}
 \mathfrak{m}_3(v,v,v)&=\mu(\mu'(v, v),v)\,, & \mathfrak{m}_4(v,v,v,v)&=\mu(\mu'(\mu'(v, v),v),v)+\tfrac12\mu(\mu(\mu''(v, v),v),v)\,,
\end{align}
cf. \eqref{somemexamples}. Though not obvious, the structure maps $\mathfrak{m}_n$ encode the entire information about the $A_\infty$-algebra.\footnote{This is not an invariant statement, of course. They encode all the structure maps as long as the $A_\infty$-algebra  is defined as above \cite{Sharapov:2018kjz}. Any diffeomorphism can change that.}

The main new result is that there is a similar flow equation for invariants. It allows one to evolve an invariant of the initial $A_\infty$-algebra according to the flow \eqref{floweq} of the $A_\infty$-algebra itself. The flow reads
\begin{align}\label{flowInv}
    \dot{S} (v,...,v)&=  S' (\mathfrak{m}(v,...,v),v,...,v) +S (\mathfrak{m}'(v,...,v),v,...,v)\,.
\end{align}
The invariants we are interested in emerge in the $\up\rightarrow0$ limit, similarly to how the structure maps of the $A_\infty$-algebra are obtained from the flow, e.g. $\mu(v,v)|_{\up=0}=v\star v$, $\mu(\mu'(v,v),v)|_{\up=0}=\phi_1(v, v)\star v$. The flow is normalized in such a way that we need to divide by the number of arguments to get the invariant
\begin{align}
    I(v_1,...,v_n)&= \tfrac{1}{n}\mathcal{S}(v_1,...,v_n)\big|_{\nu=0}\,.
\end{align}
Let us illustrate how the flow works at the lowest order. We have $S=S^{(0)}+tS^{(1)}+...$, which starts with $S(v)\equiv I_n(v)$ for any $n>0$. For example, $S^{(0)}_1(v)=\Trace{\up}{v}$ and for $S^{(1)}$ we have
\begin{align}
    S^{(1)}(v,v)&= (\pl_\up S^{(0)})[\mathfrak{m}(v,v))]+S^{(0)}[(\pl_\up\mathfrak{m})(v,v)]=\trA_\up'(\mu(v,v))+\trA_\up(\mu'(v,v))\,,
\end{align}
and at $\up=0$ we find $I(v,v)=\tfrac12 \Trace{1}{v\star v} +\tfrac12 \Trace{0}{\phi_1(v,v)}$. By iterating the flow a few times (and for a different number of arguments) one easily convinces oneself that the solution presented in the main text is correct. The general proof can be obtained as follows. First of all, at the $n$-th order we find 
\begin{align}
    \mathfrak{m}^{(n)}(v_0,...,v_{n+1})&\equiv \sum_{|\alpha|=n} \tfrac{1}{\alpha_0!...\alpha_n!}\,\mu^{(\alpha_n)}(..., \mu^{(\alpha_0)}(v_0,v_1),...,v_{n+1})\,,
\end{align}
where $\mu^{(\alpha)}$ denotes the order $\alpha$ derivative of $\mu$ w.r.t. $\nu$. The sum, however, is not over all partitions $\{\alpha_i\}$, but only over very specific ones \cite{Sharapov:2018kjz}. To each term in the decomposition hereinafter we associate the following monomial in auxiliary variables $z_0,...,z_n$:
\begin{align}
    \tfrac{1}{\alpha_0!...\alpha_n!}\,\mu^{(\alpha_n)}(...\mu^{(\alpha_0)}(v_0,v_1),...,v_{n+1}) \Longleftrightarrow \frac{z_0^{\alpha_0}...z_n^{\alpha_n}}{\alpha_0!...\alpha_n!}\,.
\end{align}
Equivalently, in terms of the generating functions $e^{u(z_0+...+z_n)}f_n(z_0,...,z_n)\leftrightarrow \mathfrak{m}^{(n)}$ of $z$'s, the flow \eqref{floweq} implies
\begin{align}\label{flowrep}
    (n+1)f_{n+1}(z_0,...,z_{n+1})&= \sum_{i+j=n} (z_0+...+z_i)f_i(z_0,...,z_i)f_j(z_{i+1},...,z_{n+1})\,.
\end{align}
The initial data $f_0(z_0)=1$ leads to $f_1(z_0,z_1)=z_0$, $f_2(z_0,z_1,z_2)=z_0^2/2+z_0z_1$, 
\begin{align}
    f_3(z_0,z_1,z_2,z_3)&= \frac{1}{6} z_0 \left(3 z_1^2+3 z_0 z_1+z_0^2+3 z_2 \left(2 z_1+z_0\right)\right)\,, \quad \text{etc\,.}
\end{align}
Note, that $f_{n}$ does not depend on $z_{n}$. Also, $f(z_0=0,z_1,...)=0$, i.e. there is no constant term in $z_0$ (except for $f_0$ itself). Now, at the $n$-th order we expect to find terms of order $\up^0,...,\up^n$ in the expansion of $v^{\prd(n+1)}=\mu(...(\mu(v,v),v),...,v)$ in front of $\trA_n,...,\trA_0$, respectively. It is easy to see that
\begin{align}
    v^{\prd(n+1)}_k \Longleftrightarrow \frac{1}{k!} (\pl_\nu)^k\mu(...,(\mu(v,v),v),...,v) \Longleftrightarrow \frac{(z_1+...+z_n)^k}{k!}\,.
\end{align}
Now, we can express \eqref{flowInv} in terms of generating functions to find
\begin{align}\label{mainrel}
\begin{aligned}
    (n+1) \frac{(z_0+...+z_n)^k}{k!}&=
    \sum_{i=0}^{k}f_i(z_0,...,z_i)\frac{(z_{i+1}+...+z_n)^{k-i}}{(k-i)!} (n+1-i)+\\
    &+\sum_{i=0}^{k-1}f_i(z_0,...,z_i)\frac{(z_{i+1}+...+z_n)^{k-i-1}}{(k-i-1)!}(z_0+...+z_i)\,.
\end{aligned}
\end{align}
where $k\in [0,n+1]$. This is a coefficient of $\trA_{n+1-k}$ in \eqref{flowInv}, where we also substituted our conjectured solution. We have to prove that the relation above is indeed satisfied for $f_i$'s that are obtained from \eqref{flowrep}.

As a particular case of relations \eqref{mainrel}, we find for $k=n+1$ a very interesting representation of multinomials:
\begin{align}\label{multirepA}
    \frac{(z_0+...+z_n)^n}{n!}&= \sum_{i=0}^{n}f_i(z_0,...,z_i)\frac{(z_{i+1}+...+z_n)^{n-i}}{(n-i)!}\,,
\end{align}
which can be massaged to an even more general relation by appending more $z$'s:
\begin{align}\label{multirep}
    \frac{(z_0+...+z_n)^k}{k!}&= \sum_{i=0}^{k}f_i(z_0,..,z_i)\frac{(z_{i+1}+...+z_n)^{k-i}}{(k-i)!}\,.
\end{align}
Note that $n$ can be arbitrary large here, $n\geq k$, since $z_i$ with $i>k$ do not appear as arguments of $f$'s. According to this formula, the structure maps of the $A_\infty$-algebra can be understood as complements of $v^{\prd(n)}_i$ that allow one to build $v^{\prd(n+2)}_n$ out of $v^{\prd(k)}_i$ of lower degree. This gives another useful description of the $A_\infty$-algebra constructed in \cite{Sharapov:2018kjz}. We can also describe $f_i$ explicitly as 
\begin{align}\label{onerep}
    f_i(z_0,..,z_i)&= \sum_{\substack{\alpha_0+...+\alpha_{i-1}=i \\
    \alpha_0+...+\alpha_j\neq j,\,\, j\in [0,i-2]}} \frac{z_0^{\alpha_0}...z_{i-1}^{\alpha_{i-1}}}{\alpha_0!...\alpha_{i-1}!}\,,
\end{align}
which follows directly from expanding $(z_0+...+z_n)^k$ and picking terms at a given $f(z_0,..,z_i)(z_{i+1}+...+z_n)^{k-i}$. Note that \eqref{onerep} does not depend on $k$, which explains universality of $f$'s. An equivalent form reads
\begin{align}
    f_i(z_0,..,z_i)&= \sum_{\substack{\alpha_0+...+\alpha_{i}=i \\
    \alpha_0+...+\alpha_j\neq j,\,\, j\in [0,i-1]}} \frac{z_0^{\alpha_0}...z_{i}^{\alpha_{i}}}{\alpha_0!...\alpha_{i}!}\,.
\end{align}
It follows that $|\alpha_0+...+\alpha_j|>j$ for $j\in[0,n-1]$. Therefore, $\alpha_i=0$  and the function does not depend on the last argument. Plotting $|\alpha_0+...+\alpha_j|$ against $j$ the graph has to be above the diagonal for $j<i$ and meets it at $j=i$. This also explains the equality  $f_i(z_0=0,...)=0$ (except for $f_0\equiv1$). 

One can show that \eqref{mainrel} is true for all admissible $k$ provided $f_i$ satisfy \eqref{multirepA} or \eqref{multirep}, which amounts to
\begin{align}
\begin{aligned}
    0&=
    \sum_{i=0}^{k}f_i(z_0,...,z_i)\left((-i)\frac{(z_{i+1}+...+z_n)^{k-i}}{(k-i)!}+\frac{(z_{i+1}+...+z_n)^{k-i-1}}{(k-i-1)!}(z_0+...+z_i)\right)\,.
\end{aligned}
\end{align}
This follows after splitting $z_0+..+z_i=(z_0+..z_n)-(z_{i+1}+...+z_n)$ and combining the terms to make \eqref{multirepA} appear.

It remains to prove that \eqref{flowrep} and \eqref{multirep} have exactly the same solutions provided we plug in the same initial data $f_0=1$. The explicit form \eqref{onerep} is not very convenient to achieve that. Instead, we prove that the functions obey the same differential equation. Starting from \eqref{flowrep} we see that
\begin{align}
    \pl_{z_0}f_{n+1}(z_0,...,z_{n+1})&= \sum_{i+j=n} f_i(z_0,...,z_i)f_j(z_{i+1},...,z_{n+1})\,,
\end{align}
which can be proved by induction. Next, with the help of the same induction and the relation here-above we get
\begin{align}\label{flowrepMod}
    \pl_{z_0}f_{n+1}(z_0,...,z_{n+1})&= f_{n}(z_0+z_1,z_2,...,z_{n+1})\,.
\end{align}
Given that there is no $z_0$-constant term for $n>0$, this equation uniquely reconstructs $f_{n+1}$ starting from $f_n$. Now we would like to show that \eqref{flowrepMod} is also satisfied as a consequence of \eqref{multirepA}. To do that, we first rewrite \eqref{multirepA} in a form suitable for induction
\begin{align}
    f_n(z_0,...,z_n)&= \frac{(z_0+...+z_n)^n}{n!}- \sum_{i=0}^{n-1}f_i(z_0,..,z_i)\frac{(z_{i+1}+...+z_n)^{n-i}}{(n-i)!}\,.
\end{align}
Next, we apply $\pl_{z_0}$ to both sides and recall that $f_0(z_0)=1$ and $\pl_{z_0}f_1(z_0,z_1)=1$, which results in the same \eqref{flowrepMod}. Therefore, \eqref{flowrep} and \eqref{mainrel} uniquely determine the same solution $f_i$ and are mutually consistent. Thus, the generating function \eqref{completecor} is the invariant of the deformed higher spin symmetry.

\footnotesize
\providecommand{\href}[2]{#2}\begingroup\raggedright\endgroup

\end{document}

%% file: YoungDiagrams.tex
\newcommand{\bep}{\begin{picture}}
\newcommand{\eep}{\end{picture}}

\newcounter{YoungHeight}\newcounter{YoungWidth}

\newcounter{Mul1}\newcounter{Mul2}\newcounter{Mul3}\newcounter{Mul4}
\newcounter{A0}\newcounter{A1}\newcounter{A2}
\newcounter{B3}
\newcounter{C3}\newcounter{C4}
\newcounter{D1}\newcounter{D2}\newcounter{D3}
\newcounter{T0}\newcounter{T1}

\newlength{\txtHShift}

\newlength{\txtWidth}

\newcommand{\HalfLength}[2]{\setcounter{Mul1}{#1}\setcounter{Mul2}{#1}\addtocounter{Mul1}{\value{Mul2}}\addtocounter{Mul1}{\value{Mul2}}%
\addtocounter{Mul1}{\value{Mul2}}\addtocounter{Mul1}{\value{Mul2}}\setcounter{#2}{\value{Mul1}}}

\newcommand{\Add}[3]{\setcounter{#1}{#2}\addtocounter{#1}{#3}}

\newcommand{\Length}[1]{#10}
\newcommand{\YoungScale}{}

\newcommand{\shiftedText}[2]{{\hspace{#1}#2}}
\newcommand{\calcHShift}[1]{\settowidth{\txtWidth}{#1}\setlength{\txtHShift}{-0.5\txtWidth}}

\newcommand{\TextTop}[3]{{\calcHShift{#1}\HalfLength{#2}{T0}\Add{T1}{\Length{#3}}{-9}\put(\value{T0},\value{T1}){\shiftedText{\txtHShift}{#1}}}}

\newcommand{\RectT}[3]{\bep(\Length{#1},\Length{#2})\put(0,0){\line(1,0){\Length{#1}}}\put(0,0){\line(0,1){\Length{#2}}}%
\put(\Length{#1},\Length{#2}){\line(-1,0){\Length{#1}}}\put(\Length{#1},\Length{#2}){\line(0,-1){\Length{#2}}}#3{#1}{#2}\eep}

\newcommand{\RectBRow}[4]{{\bep(\Length{#1},20)\put(0,0){\RectT{#2}{1}{\TextTop{#4}}}%
\put(0,10){\RectT{#1}{1}{\TextTop{#3}}}\eep}}

\newcommand{\BlockApar}[2]{\parbox{\Length{#1}pt}{\YoungScale\bep(\Length{#1},\Length{#2}){\Add{A1}{#1}{1}\Add{A2}{#2}{1}}%
\multiput(0,0)(10,0){\value{A1}}{\line(0,1){\Length{#2}}}\multiput(0,0)(0,10){\value{A2}}{\line(1,0){\Length{#1}}}%
\setcounter{YoungHeight}{\Length{#2}}\setcounter{YoungWidth}{\Length{#1}}\eep}}

\newcommand{\YoungpAA}{\BlockApar{1}{2}}

\newcommand{\YoungpBB}{\BlockApar{2}{2}}

\newcommand{\YoungpCC}{\BlockApar{3}{2}}






%% file: SBHSarxiv.bbl
\begin{thebibliography}{100}


\bibitem{Belavin:1984vu}
A.~A. Belavin, A.~M. Polyakov, and A.~B. Zamolodchikov, ``{Infinite Conformal
  Symmetry in Two-Dimensional Quantum Field Theory},''
  \href{http://dx.doi.org/10.1016/0550-3213(84)90052-X}{{\em Nucl. Phys. B}
  {\bfseries 241} (1984) 333--380}.

\bibitem{Beisert:2006fmy}
N.~Beisert, ``{The S-matrix of AdS / CFT and Yangian symmetry},''
  \href{http://dx.doi.org/10.22323/1.038.0002}{{\em PoS} {\bfseries SOLVAY}
  (2006) 002}, \href{http://arxiv.org/abs/0704.0400}{{\ttfamily arXiv:0704.0400
  [nlin.SI]}}.

\bibitem{Drummond:2009fd}
J.~M. Drummond, J.~M. Henn, and J.~Plefka, ``{Yangian symmetry of scattering
  amplitudes in N=4 super Yang-Mills theory},''
  \href{http://dx.doi.org/10.1088/1126-6708/2009/05/046}{{\em JHEP} {\bfseries
  05} (2009) 046}, \href{http://arxiv.org/abs/0902.2987}{{\ttfamily
  arXiv:0902.2987 [hep-th]}}.

\bibitem{Beisert:2010jr}
N.~Beisert {\em et~al.}, ``{Review of AdS/CFT Integrability: An Overview},''
  \href{http://dx.doi.org/10.1007/s11005-011-0529-2}{{\em Lett. Math. Phys.}
  {\bfseries 99} (2012) 3--32},
  \href{http://arxiv.org/abs/1012.3982}{{\ttfamily arXiv:1012.3982 [hep-th]}}.

\bibitem{Zwiebach:1992ie}
B.~Zwiebach, ``{Closed string field theory: Quantum action and the B-V master
  equation},'' {\em Nucl. Phys.} {\bfseries B390} (1993) 33--152,
\href{http://arxiv.org/abs/hep-th/9206084}{{\ttfamily arXiv:hep-th/9206084
  [hep-th]}}.

\bibitem{Gaberdiel:1997ia}
M.~R. Gaberdiel and B.~Zwiebach, ``{Tensor constructions of open string
  theories. 1: Foundations},'' {\em Nucl. Phys.} {\bfseries B505} (1997)
  569--624,
\href{http://arxiv.org/abs/hep-th/9705038}{{\ttfamily arXiv:hep-th/9705038
  [hep-th]}}.

\bibitem{Kajiura:2003ax}
H.~Kajiura, ``{Noncommutative homotopy algebras associated with open
  strings},'' {\em Rev. Math. Phys.} {\bfseries 19} (2007) 1--99,
\href{http://arxiv.org/abs/math/0306332}{{\ttfamily arXiv:math/0306332
  [math-qa]}}.

\bibitem{Lada:1992wc}
T.~Lada and J.~Stasheff, ``{Introduction to SH Lie algebras for physicists},''
  {\em Int. J. Theor. Phys.} {\bfseries 32} (1993) 1087--1104,
\href{http://arxiv.org/abs/hep-th/9209099}{{\ttfamily arXiv:hep-th/9209099
  [hep-th]}}.

\bibitem{Alexandrov:1995kv}
M.~Alexandrov, M.~Kontsevich, A.~Schwarz, and O.~Zaboronsky, ``{The Geometry of
  the Master Equation and Topological Quantum Field Theory},'' {\em Int. J.
  Mod. Phys.} {\bfseries A12} (1997) 1405--1429,
\href{http://arxiv.org/abs/hep-th/9502010}{{\ttfamily arXiv:hep-th/9502010
  [hep-th]}}.

\bibitem{Barnich:2004cr}
G.~Barnich, M.~Grigoriev, A.~Semikhatov, and I.~Tipunin, ``{Parent field theory
  and unfolding in BRST first-quantized terms},'' {\em Commun. Math. Phys.}
  {\bfseries 260} (2005) 147--181,
\href{http://arxiv.org/abs/hep-th/0406192}{{\ttfamily arXiv:hep-th/0406192
  [hep-th]}}.

\bibitem{Hohm:2017pnh}
O.~Hohm and B.~Zwiebach, ``{$L_{\infty}$ Algebras and Field Theory},'' {\em
  Fortsch. Phys.} {\bfseries 65} no.~3-4, (2017) 1700014,
\href{http://arxiv.org/abs/1701.08824}{{\ttfamily arXiv:1701.08824 [hep-th]}}.

\bibitem{Zilch}
T.~W.~B. Kibble, ``Conservation laws for free fields,''
  \href{http://dx.doi.org/10.1063/1.1704363}{{\em Journal of Mathematical
  Physics} {\bfseries 6} no.~7, (1965) 1022--1026}.

\bibitem{Deser:1980fk}
S.~Deser and H.~Nicolai, ``{Nonabelian Zilch},''
{\em Phys. Lett.} {\bfseries 98B} (1981) 45--47.

\bibitem{Maldacena:2011jn}
J.~Maldacena and A.~Zhiboedov, ``{Constraining Conformal Field Theories with A
  Higher Spin Symmetry},''
\href{http://arxiv.org/abs/1112.1016}{{\ttfamily arXiv:1112.1016 [hep-th]}}.

\bibitem{Boulanger:2013zza}
N.~Boulanger, D.~Ponomarev, E.~D. Skvortsov, and M.~Taronna, ``{On the
  uniqueness of higher-spin symmetries in AdS and CFT},''
  \href{http://dx.doi.org/10.1142/S0217751X13501625}{{\em Int. J. Mod. Phys.}
  {\bfseries A28} (2013) 1350162},
\href{http://arxiv.org/abs/1305.5180}{{\ttfamily arXiv:1305.5180 [hep-th]}}.

\bibitem{Alba:2013yda}
V.~Alba and K.~Diab, ``{Constraining conformal field theories with a higher
  spin symmetry in d=4},''
\href{http://arxiv.org/abs/1307.8092}{{\ttfamily arXiv:1307.8092 [hep-th]}}.

\bibitem{Alba:2015upa}
V.~Alba and K.~Diab, ``{Constraining conformal field theories with a higher
  spin symmetry in $d> 3$ dimensions},''
\href{http://arxiv.org/abs/1510.02535}{{\ttfamily arXiv:1510.02535 [hep-th]}}.

\bibitem{Maldacena:2012sf}
J.~Maldacena and A.~Zhiboedov, ``Constraining conformal field theories with a
  slightly broken higher spin symmetry,'' {\em Classical and Quantum Gravity}
  {\bfseries 30} no.~10, (2013) 104003.

\bibitem{Sharapov:2018kjz}
A.~Sharapov and E.~Skvortsov, ``{$A_\infty$ algebras from slightly broken
  higher spin symmetries},''
  \href{http://dx.doi.org/10.1007/JHEP09(2019)024}{{\em JHEP} {\bfseries 09}
  (2019) 024},
\href{http://arxiv.org/abs/1809.10027}{{\ttfamily arXiv:1809.10027 [hep-th]}}.

\bibitem{Sharapov:2020quq}
A.~Sharapov and E.~Skvortsov, ``{Characteristic Cohomology and Observables in
  Higher Spin Gravity},'' \href{http://dx.doi.org/10.1007/JHEP12(2020)190}{{\em
  JHEP} {\bfseries 12} (2020) 190},
\href{http://arxiv.org/abs/2006.13986}{{\ttfamily arXiv:2006.13986 [hep-th]}}.

\bibitem{Giombi:2011kc}
S.~Giombi, S.~Minwalla, S.~Prakash, S.~P. Trivedi, S.~R. Wadia, and X.~Yin,
  ``{Chern-Simons Theory with Vector Fermion Matter},'' {\em Eur. Phys. J.}
  {\bfseries C72} (2012) 2112,
\href{http://arxiv.org/abs/1110.4386}{{\ttfamily arXiv:1110.4386 [hep-th]}}.

\bibitem{Aharony:2012nh}
O.~Aharony, G.~Gur-Ari, and R.~Yacoby, ``{Correlation Functions of Large N
  Chern-Simons-Matter Theories and Bosonization in Three Dimensions},'' {\em
  JHEP} {\bfseries 12} (2012) 028,
\href{http://arxiv.org/abs/1207.4593}{{\ttfamily arXiv:1207.4593 [hep-th]}}.

\bibitem{Aharony:2015mjs}
O.~Aharony, ``{Baryons, monopoles and dualities in Chern-Simons-matter
  theories},'' {\em JHEP} {\bfseries 02} (2016) 093,
\href{http://arxiv.org/abs/1512.00161}{{\ttfamily arXiv:1512.00161 [hep-th]}}.

\bibitem{Karch:2016sxi}
A.~Karch and D.~Tong, ``{Particle-Vortex Duality from 3d Bosonization},'' {\em
  Phys. Rev.} {\bfseries X6} no.~3, (2016) 031043,
\href{http://arxiv.org/abs/1606.01893}{{\ttfamily arXiv:1606.01893 [hep-th]}}.

\bibitem{Seiberg:2016gmd}
N.~Seiberg, T.~Senthil, C.~Wang, and E.~Witten, ``{A Duality Web in 2+1
  Dimensions and Condensed Matter Physics},'' {\em Annals Phys.} {\bfseries
  374} (2016) 395--433,
\href{http://arxiv.org/abs/1606.01989}{{\ttfamily arXiv:1606.01989 [hep-th]}}.

\bibitem{Colombo:2012jx}
N.~Colombo and P.~Sundell, ``{Higher Spin Gravity Amplitudes From Zero-form
  Charges},''
\href{http://arxiv.org/abs/1208.3880}{{\ttfamily arXiv:1208.3880 [hep-th]}}.

\bibitem{Didenko:2012tv}
V.~Didenko and E.~Skvortsov, ``{Exact higher-spin symmetry in CFT: all
  correlators in unbroken Vasiliev theory},'' {\em JHEP} {\bfseries 1304}
  (2013) 158,
\href{http://arxiv.org/abs/1210.7963}{{\ttfamily arXiv:1210.7963 [hep-th]}}.

\bibitem{Didenko:2013bj}
V.~E. Didenko, J.~Mei, and E.~D. Skvortsov, ``{Exact higher-spin symmetry in
  CFT: free fermion correlators from Vasiliev Theory},'' {\em Phys. Rev.}
  {\bfseries D88} (2013) 046011,
\href{http://arxiv.org/abs/1301.4166}{{\ttfamily arXiv:1301.4166 [hep-th]}}.

\bibitem{Bonezzi:2017vha}
R.~Bonezzi, N.~Boulanger, D.~De~Filippi, and P.~Sundell, ``{Noncommutative
  Wilson lines in higher-spin theory and correlation functions of conserved
  currents for free conformal fields},'' {\em J. Phys.} {\bfseries A50} no.~47,
  (2017) 475401,
\href{http://arxiv.org/abs/1705.03928}{{\ttfamily arXiv:1705.03928 [hep-th]}}.

\bibitem{Giombi:2016zwa}
S.~Giombi, V.~Gurucharan, V.~Kirilin, S.~Prakash, and E.~Skvortsov, ``{On the
  Higher-Spin Spectrum in Large N Chern-Simons Vector Models},'' {\em JHEP}
  {\bfseries 01} (2017) 058,
\href{http://arxiv.org/abs/1610.08472}{{\ttfamily arXiv:1610.08472 [hep-th]}}.

\bibitem{Jain:2020puw}
S.~Jain, R.~R. John, and V.~Malvimat, ``{Constraining momentum space
  correlators using slightly broken higher spin symmetry},''
  \href{http://dx.doi.org/10.1007/JHEP04(2021)231}{{\em JHEP} {\bfseries 04}
  (2021) 231}, \href{http://arxiv.org/abs/2008.08610}{{\ttfamily
  arXiv:2008.08610 [hep-th]}}.

\bibitem{Skvortsov:2018uru}
E.~Skvortsov, ``{Light-Front Bootstrap for Chern-Simons Matter Theories},''
  {\em JHEP} {\bfseries 06} (2019) 058,
\href{http://arxiv.org/abs/1811.12333}{{\ttfamily arXiv:1811.12333 [hep-th]}}.

\bibitem{Craigie:1983fb}
N.~S. Craigie, V.~K. Dobrev, and I.~T. Todorov, ``{Conformally Covariant
  Composite Operators in Quantum Chromodynamics},''
{\em Annals Phys.} {\bfseries 159} (1985) 411--444.

\bibitem{Anselmi:1999bb}
D.~Anselmi, ``{Higher spin current multiplets in operator product
  expansions},'' {\em Class. Quant. Grav.} {\bfseries 17} (2000) 1383--1400,
\href{http://arxiv.org/abs/hep-th/9906167}{{\ttfamily arXiv:hep-th/9906167
  [hep-th]}}.

\bibitem{Alkalaev:2012rg}
K.~Alkalaev, ``{Mixed-symmetry tensor conserved currents and AdS/CFT
  correspondence},'' {\em J. Phys.} {\bfseries A46} (2013) 214007,
\href{http://arxiv.org/abs/1207.1079}{{\ttfamily arXiv:1207.1079 [hep-th]}}.

\bibitem{Walker:1970un}
M.~Walker and R.~Penrose, ``{On quadratic first integrals of the geodesic
  equations for type [22] spacetimes},''
  \href{http://dx.doi.org/10.1007/BF01649445}{{\em Commun. Math. Phys.}
  {\bfseries 18} (1970) 265--274}.

\bibitem{Nikitin1991}
A.~G. Nikitin, ``Generalized killing tensors of arbitrary rank and order,''
  \href{http://dx.doi.org/10.1007/BF01058941}{{\em Ukrainian Mathematical
  Journal} {\bfseries 43} no.~6, (Jun, 1991) 734--743}.

\bibitem{Eastwood:2002su}
M.~G. Eastwood, ``{Higher symmetries of the Laplacian},'' {\em Annals Math.}
  {\bfseries 161} (2005) 1645--1665,
\href{http://arxiv.org/abs/hep-th/0206233}{{\ttfamily arXiv:hep-th/0206233
  [hep-th]}}.

\bibitem{Konstein:2000bi}
S.~E. Konstein, M.~A. Vasiliev, and V.~N. Zaikin, ``{Conformal higher spin
  currents in any dimension and AdS / CFT correspondence},''
  \href{http://dx.doi.org/10.1088/1126-6708/2000/12/018}{{\em JHEP} {\bfseries
  12} (2000) 018}, \href{http://arxiv.org/abs/hep-th/0010239}{{\ttfamily
  arXiv:hep-th/0010239}}.

\bibitem{Dirac:1963ta}
P.~A.~M. Dirac, ``{A Remarkable representation of the 3 + 2 de Sitter group},''
{\em J. Math. Phys.} {\bfseries 4} (1963) 901--909.

\bibitem{Gunaydin:1981yq}
M.~G{\"u}naydin and C.~Saclioglu, ``{Oscillator Like Unitary Representations of
  Noncompact Groups With a Jordan Structure and the Noncompact Groups of
  Supergravity},''
{\em Commun. Math. Phys.} {\bfseries 87} (1982) 159.

\bibitem{Gunaydin:1983yj}
M.~G{\"u}naydin, ``{Oscillator like unitary representations of noncompact
  groups and supergroups and extended supergravity theories},'' in {\em {Group
  Theoretical Methods in Physics. Proceedings, 11th International Colloquium,
  Istanbul, Turkey, August 23-28, 1982}}, pp.~192--213.
\newblock
1983.
\newblock

\bibitem{Fradkin:1986ka}
E.~S. Fradkin and M.~A. Vasiliev, ``Candidate to the role of higher spin
  symmetry,''
{\em Ann. Phys.} {\bfseries 177} (1987) 63.

\bibitem{Vasiliev:1986qx}
M.~A. Vasiliev, ``Extended higher spin superalgebras and their realizations in
  terms of quantum operators,''
{\em Fortsch. Phys.} {\bfseries 36} (1988) 33--62.

\bibitem{Gunaydin:1989um}
M.~G{\"u}naydin, ``{Singleton and doubleton supermultiplets of space-time
  supergroups and infinite spin superalgebras},'' in {\em {Trieste Conference
  on Supermembranes and Physics in 2+1 Dimensions Trieste, Italy, July 17-21,
  1989}}, pp.~0442--456.
\newblock
1989.
\newblock

\bibitem{Michel}
J.-P. Michel, ``Higher symmetries of the laplacian via~quantization,''
  \href{http://dx.doi.org/10.5802/aif.2891}{{\em Annales de l'Institut Fourier}
  {\bfseries 64} no.~4, (2014) 1581--1609}.

\bibitem{Joung:2014qya}
E.~Joung and K.~Mkrtchyan, ``{Notes on higher-spin algebras: minimal
  representations and structure constants},'' {\em JHEP} {\bfseries 05} (2014)
  103,
\href{http://arxiv.org/abs/1401.7977}{{\ttfamily arXiv:1401.7977 [hep-th]}}.

\bibitem{Gunaydin:2016bqx}
M.~Gunaydin,
  \href{http://dx.doi.org/10.1142/9789813144101_0010}{``{Quasiconformal Group
  Approach to Higher Spin Algebras, their Deformations and Supersymmetric
  Extensions},''} in {\em {International Workshop on Higher Spin Gauge
  Theories}}.
\newblock 3, 2016.
\newblock \href{http://arxiv.org/abs/1603.02359}{{\ttfamily arXiv:1603.02359
  [hep-th]}}.

\bibitem{deMelloKoch:2018ivk}
R.~de~Mello~Koch, A.~Jevicki, K.~Suzuki, and J.~Yoon, ``{AdS Maps and Diagrams
  of Bi-local Holography},'' {\em JHEP} {\bfseries 03} (2019) 133,
\href{http://arxiv.org/abs/1810.02332}{{\ttfamily arXiv:1810.02332 [hep-th]}}.

\bibitem{Aharony:2020omh}
O.~Aharony, S.~M. Chester, and E.~Y. Urbach, ``{A Derivation of AdS/CFT for
  Vector Models},''
\href{http://arxiv.org/abs/2011.06328}{{\ttfamily arXiv:2011.06328 [hep-th]}}.

\bibitem{David:2020ptn}
A.~David and Y.~Neiman, ``{Higher-spin symmetry vs. boundary locality, and a
  rehabilitation of dS/CFT},''
  \href{http://dx.doi.org/10.1007/JHEP10(2020)127}{{\em JHEP} {\bfseries 10}
  (2020) 127}, \href{http://arxiv.org/abs/2006.15813}{{\ttfamily
  arXiv:2006.15813 [hep-th]}}.

\bibitem{Iazeolla:2008ix}
C.~Iazeolla and P.~Sundell, ``{A Fiber Approach to Harmonic Analysis of
  Unfolded Higher- Spin Field Equations},'' {\em JHEP} {\bfseries 10} (2008)
  022,
\href{http://arxiv.org/abs/0806.1942}{{\ttfamily arXiv:0806.1942 [hep-th]}}.

\bibitem{Giombi:2010vg}
S.~Giombi and X.~Yin, ``{Higher Spins in AdS and Twistorial Holography},'' {\em
  JHEP} {\bfseries 1104} (2011) 086,
\href{http://arxiv.org/abs/1004.3736}{{\ttfamily arXiv:1004.3736 [hep-th]}}.

\bibitem{Sleight:2016dba}
C.~Sleight and M.~Taronna, ``{Higher Spin Interactions from Conformal Field
  Theory: The Complete Cubic Couplings},'' {\em Phys. Rev. Lett.} {\bfseries
  116} no.~18, (2016) 181602,
\href{http://arxiv.org/abs/1603.00022}{{\ttfamily arXiv:1603.00022 [hep-th]}}.

\bibitem{Luders}
G.~Luders, ``{Vertauschungsrelationen zwischen verschiedenen Feldern},'' {\em
  Z. Naturforsch} {\bfseries 13A} (1958) 254.

\bibitem{Druehl:1970fz}
K.~Druehl, R.~Haag, and J.~E. Roberts, ``{On parastatistics},''
  \href{http://dx.doi.org/10.1007/BF01649433}{{\em Commun. Math. Phys.}
  {\bfseries 18} (1970) 204--226}.

\bibitem{Schmutz}
M.~Schmutz, ``Simplified bose description of para‐bose operators,''
  \href{http://dx.doi.org/10.1063/1.524614}{{\em Journal of Mathematical
  Physics} {\bfseries 21} no.~7, (1980) 1665--1666}.

\bibitem{Ohnuki1982}
Y.~Ohnuki and S.~Kamefuchi, {\em Quantum field theory and parastatistics}.
\newblock Springer, Germany, 1982.

\bibitem{Ohnuki:1984gz}
Y.~Ohnuki and S.~Kamefuchi, ``{Fermi-bose Similarity, Supersymmetry and
  Generalized Numbers. 3. Klein Transformations},''
  \href{http://dx.doi.org/10.1007/BF02902602}{{\em Nuovo Cim. A} {\bfseries 83}
  (1984) 275}.

\bibitem{Wigner1950}
E.~P. Wigner, ``Do the equations of motion determine the quantum mechanical
  commutation relations?,'' {\em Phys. Rev.} {\bfseries 77} (Mar, 1950)
  711--712.

\bibitem{Yang:1951pyq}
L.~M. Yang, ``{A Note on the Quantum Rule of the Harmonic Oscillator},''
{\em Phys. Rev.} {\bfseries 84} no.~4, (1951) 788.

\bibitem{Boulware1963}
D.~G. Boulware and S.~Deser, ``'ambiguity' of harmonic-oscillator commutation
  relations,'' \href{http://dx.doi.org/10.1007/BF02750763}{{\em Il Nuovo
  Cimento (1955-1965)} {\bfseries 30} no.~1, (Oct, 1963) 230--234}.
  \url{https://doi.org/10.1007/BF02750763}.

\bibitem{Gruber}
B.~Gruber and L.~O'Raifeartaigh, ``Uniqueness of the harmonic oscillator
  commutation relation,'' {\em Proceedings of the Royal Irish Academy. Section
  A: Mathematical and Physical Sciences} {\bfseries 63} (1963) 69--73.

\bibitem{Mukunda:1980fv}
N.~Mukunda, E.~C.~G. Sudarshan, J.~K. Sharma, and C.~L. Mehta,
  ``{Representations and properties of parabose oscillator operators. I. Energy
  position and momentum eigenstates},''
  \href{http://dx.doi.org/10.1063/1.524695}{{\em J. Math. Phys.} {\bfseries 21}
  (1980) 2386--2394}.

\bibitem{Engquist:2008mc}
J.~Engquist, ``{Anyons, Deformed Oscillator Algebras and Projectors},''
  \href{http://dx.doi.org/10.1016/j.nuclphysb.2009.02.001}{{\em Nucl. Phys. B}
  {\bfseries 816} (2009) 356--375},
  \href{http://arxiv.org/abs/0809.3226}{{\ttfamily arXiv:0809.3226 [hep-th]}}.

\bibitem{Gandhi:2021gwn}
Y.~Gandhi, S.~Jain, and R.~R. John, ``{Anyonic correlation functions in
  Chern-Simons matter theories},''
  \href{http://arxiv.org/abs/2106.09043}{{\ttfamily arXiv:2106.09043
  [hep-th]}}.

\bibitem{Metsaev:1991mt}
R.~R. Metsaev, ``{Poincare invariant dynamics of massless higher spins: Fourth
  order analysis on mass shell},''
{\em Mod. Phys. Lett.} {\bfseries A6} (1991) 359--367.

\bibitem{Metsaev:1991nb}
R.~R. Metsaev, ``{$S$ matrix approach to massless higher spins theory. 2: The
  Case of internal symmetry},''
{\em Mod. Phys. Lett.} {\bfseries A6} (1991) 2411--2421.

\bibitem{Ponomarev:2016lrm}
D.~Ponomarev and E.~D. Skvortsov, ``{Light-Front Higher-Spin Theories in Flat
  Space},'' {\em J. Phys.} {\bfseries A50} no.~9, (2017) 095401,
\href{http://arxiv.org/abs/1609.04655}{{\ttfamily arXiv:1609.04655 [hep-th]}}.

\bibitem{Skvortsov:2020wtf}
E.~Skvortsov, T.~Tran, and M.~Tsulaia, ``{More on Quantum Chiral Higher Spin
  Gravity},'' {\em Phys. Rev.} {\bfseries D101} no.~10, (2020) 106001,
\href{http://arxiv.org/abs/2002.08487}{{\ttfamily arXiv:2002.08487 [hep-th]}}.

\bibitem{Skvortsov:2020gpn}
E.~Skvortsov and T.~Tran, ``{One-loop Finiteness of Chiral Higher Spin
  Gravity},''
\href{http://arxiv.org/abs/2004.10797}{{\ttfamily arXiv:2004.10797 [hep-th]}}.

\bibitem{GurAri:2012is}
G.~Gur-Ari and R.~Yacoby, ``{Correlators of Large N Fermionic Chern-Simons
  Vector Models},'' {\em JHEP} {\bfseries 02} (2013) 150,
\href{http://arxiv.org/abs/1211.1866}{{\ttfamily arXiv:1211.1866 [hep-th]}}.

\bibitem{Li:2019twz}
Z.~Li, ``{Bootstrapping conformal four-point correlators with slightly broken
  higher spin symmetry and $3D$ bosonization},''
  \href{http://dx.doi.org/10.1007/JHEP10(2020)007}{{\em JHEP} {\bfseries 10}
  (2020) 007}, \href{http://arxiv.org/abs/1906.05834}{{\ttfamily
  arXiv:1906.05834 [hep-th]}}.

\bibitem{Kalloor:2019xjb}
R.~R. Kalloor, ``{Four-point functions in large $N$ Chern-Simons fermionic
  theories},'' \href{http://dx.doi.org/10.1007/JHEP10(2020)028}{{\em JHEP}
  {\bfseries 10} (2020) 028}, \href{http://arxiv.org/abs/1910.14617}{{\ttfamily
  arXiv:1910.14617 [hep-th]}}.

\bibitem{Turiaci:2018nua}
G.~J. Turiaci and A.~Zhiboedov, ``{Veneziano Amplitude of Vasiliev Theory},''
  \href{http://dx.doi.org/10.1007/JHEP10(2018)034}{{\em JHEP} {\bfseries 10}
  (2018) 034}, \href{http://arxiv.org/abs/1802.04390}{{\ttfamily
  arXiv:1802.04390 [hep-th]}}.

\bibitem{Jain:2021gwa}
S.~Jain and R.~R. John, ``{Relation between parity-even and parity-odd CFT
  correlation functions in three dimensions},''
  \href{http://arxiv.org/abs/2107.00695}{{\ttfamily arXiv:2107.00695
  [hep-th]}}.

\bibitem{Silva:2021ece}
J.~A. Silva, ``{Four point functions in CFT\textquoteright{}s with slightly
  broken higher spin symmetry},''
  \href{http://dx.doi.org/10.1007/JHEP05(2021)097}{{\em JHEP} {\bfseries 05}
  (2021) 097}, \href{http://arxiv.org/abs/2103.00275}{{\ttfamily
  arXiv:2103.00275 [hep-th]}}.

\bibitem{Anselmi:1998ms}
D.~Anselmi, ``{The N=4 quantum conformal algebra},'' {\em Nucl. Phys.}
  {\bfseries B541} (1999) 369--385,
\href{http://arxiv.org/abs/hep-th/9809192}{{\ttfamily arXiv:hep-th/9809192
  [hep-th]}}.

\bibitem{Skvortsov:2015pea}
E.~D. Skvortsov, ``{On (Un)Broken Higher-Spin Symmetry in Vector Models},''
\href{http://arxiv.org/abs/hep-th:1512.05994}{{\ttfamily
  arXiv:hep-th:1512.05994 [hep-th]}}.

\bibitem{Giombi:2016hkj}
S.~Giombi and V.~Kirilin, ``{Anomalous dimensions in CFT with weakly broken
  higher spin symmetry},'' {\em JHEP} {\bfseries 11} (2016) 068,
\href{http://arxiv.org/abs/1601.01310}{{\ttfamily arXiv:1601.01310 [hep-th]}}.

\bibitem{Giombi:2017rhm}
S.~Giombi, V.~Kirilin, and E.~Skvortsov, ``{Notes on Spinning Operators in
  Fermionic CFT},'' {\em JHEP} {\bfseries 05} (2017) 041,
\href{http://arxiv.org/abs/1701.06997}{{\ttfamily arXiv:1701.06997 [hep-th]}}.

\bibitem{Manashov:2016uam}
A.~N. Manashov and E.~D. Skvortsov, ``{Higher-spin currents in the Gross-Neveu
  model at 1/n$^{2}$},'' \href{http://dx.doi.org/10.1007/JHEP01(2017)132}{{\em
  JHEP} {\bfseries 01} (2017) 132},
  \href{http://arxiv.org/abs/1610.06938}{{\ttfamily arXiv:1610.06938
  [hep-th]}}.

\bibitem{Manashov:2017xtt}
A.~N. Manashov, E.~D. Skvortsov, and M.~Strohmaier, ``{Higher spin currents in
  the critical $O(N$) vector model at $1/N^{2}$},''
  \href{http://dx.doi.org/10.1007/JHEP08(2017)106}{{\em JHEP} {\bfseries 08}
  (2017) 106}, \href{http://arxiv.org/abs/1706.09256}{{\ttfamily
  arXiv:1706.09256 [hep-th]}}.

\bibitem{Komargodski:2012ek}
Z.~Komargodski and A.~Zhiboedov, ``{Convexity and Liberation at Large Spin},''
  {\em JHEP} {\bfseries 11} (2013) 140,
\href{http://arxiv.org/abs/1212.4103}{{\ttfamily arXiv:1212.4103 [hep-th]}}.

\bibitem{Didenko:2012vh}
V.~E. Didenko and E.~D. Skvortsov, ``{Towards higher-spin holography in ambient
  space of any dimension},'' {\em J. Phys.} {\bfseries A46} (2013) 214010,
\href{http://arxiv.org/abs/1207.6786}{{\ttfamily arXiv:1207.6786 [hep-th]}}.

\bibitem{Sezgin:2002rt}
E.~Sezgin and P.~Sundell, ``{Massless higher spins and holography},'' {\em
  Nucl.Phys.} {\bfseries B644} (2002) 303--370,
\href{http://arxiv.org/abs/hep-th/0205131}{{\ttfamily arXiv:hep-th/0205131
  [hep-th]}}.

\bibitem{Klebanov:2002ja}
I.~R. Klebanov and A.~M. Polyakov, ``{AdS dual of the critical $O(N)$ vector
  model},'' {\em Phys. Lett.} {\bfseries B550} (2002) 213--219,
\href{http://arxiv.org/abs/hep-th/0210114}{{\ttfamily arXiv:hep-th/0210114}}.

\bibitem{Sezgin:2003pt}
E.~Sezgin and P.~Sundell, ``{Holography in 4D (super) higher spin theories and
  a test via cubic scalar couplings},'' {\em JHEP} {\bfseries 0507} (2005) 044,
\href{http://arxiv.org/abs/hep-th/0305040}{{\ttfamily arXiv:hep-th/0305040
  [hep-th]}}.

\bibitem{Leigh:2003gk}
R.~G. Leigh and A.~C. Petkou, ``{Holography of the N=1 higher spin theory on
  AdS(4)},'' {\em JHEP} {\bfseries 0306} (2003) 011,
\href{http://arxiv.org/abs/hep-th/0304217}{{\ttfamily arXiv:hep-th/0304217
  [hep-th]}}.

\bibitem{Bekaert:2015tva}
X.~Bekaert, J.~Erdmenger, D.~Ponomarev, and C.~Sleight, ``{Quartic AdS
  Interactions in Higher-Spin Gravity from Conformal Field Theory},'' {\em
  JHEP} {\bfseries 11} (2015) 149,
\href{http://arxiv.org/abs/1508.04292}{{\ttfamily arXiv:1508.04292 [hep-th]}}.

\bibitem{Sleight:2017pcz}
C.~Sleight and M.~Taronna, ``{Higher-Spin Gauge Theories and Bulk Locality},''
  {\em Phys. Rev. Lett.} {\bfseries 121} no.~17, (2018) 171604,
\href{http://arxiv.org/abs/1704.07859}{{\ttfamily arXiv:1704.07859 [hep-th]}}.

\bibitem{Ponomarev:2017qab}
D.~Ponomarev, ``{A Note on (Non)-Locality in Holographic Higher Spin
  Theories},'' {\em Universe} {\bfseries 4} no.~1, (2018) 2,
\href{http://arxiv.org/abs/1710.00403}{{\ttfamily arXiv:1710.00403 [hep-th]}}.

\bibitem{Giombi:2009wh}
S.~Giombi and X.~Yin, ``{Higher Spin Gauge Theory and Holography: The
  Three-Point Functions},'' {\em JHEP} {\bfseries 1009} (2010) 115,
\href{http://arxiv.org/abs/0912.3462}{{\ttfamily arXiv:0912.3462 [hep-th]}}.

\bibitem{Boulanger:2015ova}
N.~Boulanger, P.~Kessel, E.~D. Skvortsov, and M.~Taronna, ``{Higher spin
  interactions in four-dimensions: Vasiliev versus Fronsdal},'' {\em J. Phys.}
  {\bfseries A49} no.~9, (2016) 095402,
\href{http://arxiv.org/abs/1508.04139}{{\ttfamily arXiv:1508.04139 [hep-th]}}.

\bibitem{Skvortsov:2015lja}
E.~D. Skvortsov and M.~Taronna, ``{On Locality, Holography and Unfolding},''
  {\em JHEP} {\bfseries 11} (2015) 044,
\href{http://arxiv.org/abs/1508.04764}{{\ttfamily arXiv:1508.04764 [hep-th]}}.

\bibitem{Vasiliev:1988sa}
M.~A. Vasiliev, ``Consistent equations for interacting massless fields of all
  spins in the first order in curvatures,''
{\em Annals Phys.} {\bfseries 190} (1989) 59--106.

\bibitem{Vasiliev:1990cm}
M.~A. Vasiliev, ``{Closed equations for interacting gauge fields of all
  spins},''
{\em JETP Lett.} {\bfseries 51} (1990) 503--507.

\bibitem{Sezgin:2005pv}
E.~Sezgin and P.~Sundell, ``{An Exact solution of 4-D higher-spin gauge
  theory},'' {\em Nucl.Phys.} {\bfseries B762} (2007) 1--37,
\href{http://arxiv.org/abs/hep-th/0508158}{{\ttfamily arXiv:hep-th/0508158
  [hep-th]}}.

\bibitem{Engquist:2005yt}
J.~Engquist and P.~Sundell, ``{Brane partons and singleton strings},'' {\em
  Nucl. Phys.} {\bfseries B752} (2006) 206--279,
\href{http://arxiv.org/abs/hep-th/0508124}{{\ttfamily arXiv:hep-th/0508124
  [hep-th]}}.

\bibitem{Colombo:2010fu}
N.~Colombo and P.~Sundell, ``{Twistor space observables and quasi-amplitudes in
  4D higher spin gravity},'' {\em JHEP} {\bfseries 1111} (2011) 042,
\href{http://arxiv.org/abs/1012.0813}{{\ttfamily arXiv:1012.0813 [hep-th]}}.

\bibitem{Sezgin:2011hq}
E.~Sezgin and P.~Sundell, ``{Geometry and Observables in Vasiliev's Higher Spin
  Gravity},'' {\em JHEP} {\bfseries 07} (2012) 121,
\href{http://arxiv.org/abs/1103.2360}{{\ttfamily arXiv:1103.2360 [hep-th]}}.

\bibitem{Vasiliev:2015mka}
M.~A. Vasiliev, ``{Invariant Functionals in Higher-Spin Theory},''
  \href{http://dx.doi.org/10.1016/j.nuclphysb.2017.01.001}{{\em Nucl. Phys. B}
  {\bfseries 916} (2017) 219--253},
  \href{http://arxiv.org/abs/1504.07289}{{\ttfamily arXiv:1504.07289
  [hep-th]}}.

\bibitem{Sharapov:2021drr}
A.~Sharapov and E.~Skvortsov, ``{Higher Spin Gravities and Presymplectic AKSZ
  Models},''
\href{http://arxiv.org/abs/2102.02253}{{\ttfamily arXiv:2102.02253 [hep-th]}}.

\bibitem{Neiman:2017mel}
Y.~Neiman, ``{The holographic dual of the Penrose transform},'' {\em JHEP}
  {\bfseries 01} (2018) 100,
\href{http://arxiv.org/abs/1709.08050}{{\ttfamily arXiv:1709.08050 [hep-th]}}.

\bibitem{Neiman:2018ufb}
Y.~Neiman, ``{Holographic quantization of linearized higher-spin gravity in the
  de Sitter causal patch},'' {\em JHEP} {\bfseries 11} (2018) 033,
\href{http://arxiv.org/abs/1809.07270}{{\ttfamily arXiv:1809.07270 [hep-th]}}.

\bibitem{Cuntz}
J.~Cuntz, G.~Skandalis, and B.~Tsygan, {\em {Cyclic Homology in Non-Commutative
  Geometry}}.
\newblock Springer, 2004.

\bibitem{stasheff2018linfty}
J.~Stasheff, ``{$L_\infty$ and $A_\infty$-structures: then and now},''
  \href{http://arxiv.org/abs/1809.02526}{{\ttfamily {arXiv}:1809.02526
  [{math.QA}]}}.

\bibitem{Getzler93}
E.~Getzler, ``{Cartan homotopy formulas and the Gauss--Manin connection in
  cyclic homology},'' in {\em Quantum deformations of algebras and their
  representations}, vol.~7, pp.~65--78.
\newblock Bar-Ilan University, Ramat-Gan, 1993.

\end{thebibliography}
